\documentclass[journal]{IEEEtran}

\IEEEoverridecommandlockouts

\usepackage{cite}
\usepackage{amsmath,amssymb,amsfonts}
\usepackage{algorithmic}
\usepackage{graphicx}
\usepackage{textcomp}
\usepackage[utf8]{inputenc}
\usepackage{textgreek}
\usepackage{color}
\usepackage{url}
\usepackage{dblfloatfix}
\usepackage{multirow}
\usepackage{float}
\usepackage{subfigure}
\usepackage{textcomp}
\usepackage{enumerate}
\usepackage{hyperref,cleveref}

\makeatletter
   \def\hrulefill{\leavevmode\leaders\hrule height 1pt\hfill\kern\z@}
\makeatother
\usepackage{wrapfig}
\usepackage{listings}

\definecolor{gris245}{RGB}{245,245,245}
\definecolor{olive}{RGB}{50,140,50}
\definecolor{brun}{RGB}{175,100,80}
\usepackage{fancybox}
\usepackage{textcase}
\usepackage{caption}

\usepackage{xcolor}
\usepackage[linesnumbered,ruled,vlined]{algorithm2e}

\SetCommentSty{mycommfont}

\SetKwInput{KwInput}{Input}                
\SetKwInput{KwOutput}{Output}              

\makeatletter
\newenvironment{CenteredBox}{%
\begin{Sbox}}{
\end{Sbox}\centerline{\parbox{\wd\@Sbox}{\TheSbox}}}
\makeatother
\begin{document}

\title{Highly Accurate Closed-form Approximation for the \\ Probability of Detection of Weibull Fluctuating
Targets in Non-Coherent Detectors}

\author{Fernando Darío Almeida García, Andrea Carolina Flores Rodriguez and Gustavo Fraidenraich \thanks{F.~D.~A.~García and G.~Fraidenraich are with the Wireless Technology Laboratory, Department of Communications, School of Electrical and Computer Engineering, University of Campinas, 13083-852 Campinas, SP, Brazil, Tel.: +55 (19) 3788-5106, E-mails: $\{\text{ferdaral,gf}\}$@decom.fee.unicamp.br. } 
\thanks{A. C. F. Rodriguez is with EMBRAER, Campinas, Brazil, Tel.: +55 19 2101-8800, E-mail: andrea.rodriguez@embraer.com.br.}
} 

\maketitle

\begin{abstract}
In this paper, we derive a highly accurate approximation for the probability of detection (PD) of a non-coherent detector operating with Weibull fluctuation targets.
To do so, we assume a pulse--to--pulse decorrelation during the coherent processing interval (CPI).
Specifically, the proposed approximation is given in terms of: \textbf{i)} a closed-form expression derived in terms of the Fox's $H$-function, for which we also provide a portable and efficient MATHEMATICA routine; and \textbf{ii)} a fast converging series obtained through a comprehensive calculus of residues.
Both solutions are fast and provide very accurate results.
In particular, our series representation, besides being a more tractable solution, also exhibits impressive savings in computational load and computation time compared to previous studies.
Numerical results and Monte-Carlo simulations corroborated the validity of our expressions.
\end{abstract}

\begin{IEEEkeywords}
Probability of detection, non-coherent detector, Weibull fluctuating targets, Fox's  $H$-function.
\end{IEEEkeywords}

\vspace{-0.05cm}
\section{Introduction}
The target's radar cross section (RCS) plays an important role in radar detection. 
Specifically, RCS is a measure that describes the amount of energy reflected by a target and, therefore, has a direct impact on the received target echo power.
In general, RCS is a complex function of: target geometry and material composition; position of transmitter relative to target; position of receiver relative to target; frequency or wavelength; transmitter polarization; and receiver polarization~\cite{richards10}.
Since the target's RCS is extremely sensitive to the above parameters, it is common and more practical to use statistical models to capture its behavior~\cite{richards14}.
This argument leads to consider the target's RCS as a random variable (RV) with a specified probability density function (PDF).
It is important to emphasize  that using statistical models for the RCS does not imply that the actual RCS is random. If it was possible to describe the target surface shape, materials and location in enough detail, then the target's RCS could in principle be calculated accurately using deterministic approaches~\cite{Skiliman85}. However, in practice, this task seems to be extremely complicated and too demanding to be executed.

Some common statistical models for the target's RCS are the Exponential and the fourth-degree Chi-square distributions. Both distributions are part of the well-known Swerling models, also known as fluctuating target models~\cite{swerling60}. 
The Exponential distribution arises when there is a large number of individual scatterers randomly distributed in space and each with approximately the same individual RCS. 
The Exponential distribution is used in the Sweling cases I and II~\cite{swerling60,Shnidman95,Shnidman03}.
For the case when there is a large number of individual scatterers, one dominant and the rest with the same RCS, the Exponential distribution is no longer a good fit for the target's RCS. The noncentral Chi-square distribution with two degrees of freedom is the exact PDF for this case, but it is considered somewhat difficult to work with because the expression for the PDF contains a Bessel function. For this reason, the fourth-degree Chi-square distribution is used in the Swerling cases III and IV since it is a more analytically tractable approximation~\cite{Shnidman03,Shnidman98,Lim19}.

More robust target models emerge so as to accurately describe the complex behaviour of the target's RCS. Among them, we highlight the Log-normal, Chi-square and Weibull target models. These models are widely used in high-resolution radars, in which the resolution cell\footnote{The ability of a radar system to resolve two targets over range, azimuth, and elevation defines its resolution cell~\cite{richards10}.} is small enough to contain a reduced number of scatterers~\cite{barton13,mahafza13,Weiner98,Kanter86}. 
In particular, the Log-normal and Weibull target models provide an excellent empirical fit to observed data since they exhibit longer tails than common distributions. A longer tail means that there is a greater probability of observing high values of RCS.
For instance, the Weibull fluctuating model has attracted attention of many communications fields due to its applicability. For example, since the Weibull model is a two-parameter distribution, its mean and variance can be adjusted independently, thereby serving as a suitable fit for a wider range of measured data~\cite{Shnidman99,Sekine81,Schleher76}.
Moreover, the Weibull model summarizes the Exponential (in power) and Rayleigh (in voltage) target models.

Non-coherent detectors made use of the aforementioned fluctuating target models in order to obtain the system performance. This is carried out by deriving the probability of the detection (PD) from a block of $N$ independent or correlated echo samples, which are collected during a coherent processing interval (CPI)~\cite{kay93,kay98,blake86}. 
Important works have analyzed radar performance considering robust fluctuating target models. For example, in  \cite{Cui13}, the authors derived an analytical expression for the PD considering a Chi-square fluctuating target model. To do so, the authors assumed that the $N$ echo samples bear a certain degree of correlation.
In \cite{Cui14}, the authors obtained an exact expression for the PD considering the Weibull fluctuating target model, in which the $N$ echo samples were assumed to be independent of each other.
However, this expression was derived in terms of nested infinite sum-products, thereby showing a high computational burden and a high mathematical complexity that tends to grow as the number of echo samples increases.
This is mainly due to the intricate and arduous task to obtain the exact PDF of the sum of Weibull RVs (cf.~\cite{Yilmaz09conf,You16,hfox,Rahama18} for a detailed discussion on this). 
We aim to alleviate the analytical evaluation of the PD.

In this paper, capitalizing on a useful result for the sum of independent Weibull RVs \cite{Filho06}, we derive a highly accurate approximation for the PD of a non-coherent detector operating with Weibull fluctuation targets.
Specifically, the proposed approximation is given in terms of: \textbf{i)} a closed-form expression derived in terms of the Fox's $H$-function, for which we also provide a portable and efficient MATHEMATICA routine; and \textbf{ii)} a fast converging series obtained through a comprehensive calculus of residues.
Both solutions are fast and provide very accurate results, as shall be seen in Section~\ref{eq: Sample Numerical Results}.
In particular, our series representation exhibits impressive savings in computational load and computation time compared to~\cite[Eq. (30)]{Cui14}. 

The remainder of this paper is organized as follows. Section~\ref{sec: preliminaries} introduces the multivariate Fox's $H$-function.
Section~\ref{sec:System Model} presents the system model for the non-coherent detector. 
Section~\ref{sec: Sum Statistics} summarizes relevant results for the sum of independent Weibull RVs.
Section~\ref{sec: Detection Performance} analyzes the performance of non-coherent detectors considering target fluctuations.
Section~\ref{eq: Sample Numerical Results} discusses the representative numerical results.
Finally, Section~\ref{sec: Conclusions} provides some concluding remarks.

In what follows, $f_{(\cdot)}(\cdot)$ denotes PDF; $\left| \cdot \right|$, modulus of a complex
number; $\left( \cdot\right)^T$, transposition; $\mathbb{E}\left[ \cdot \right]$, expectation; $\mathbb{V}\left[ \cdot \right]$, variance; $\left( \cdot\right)^T$, transposition; $\left( \cdot\right) ^{-1}$, matrix inversion; and $\mathcal{U}\left(a,b \right)$ denotes a uniform distribution over the interval $\left[a,b \right]$.

\section{Preliminaries}
\label{sec: preliminaries}
In this section, we introduce the multivariate Fox's $H$-function, as it will be extensively exploited throughout this work.
\subsection{The Multivariate Fox H-function}
\label{subsec:The Multivariate Fox H-function}
The Fox's $H$-function has been recently used in a wide variety of applications, including mobile communications and radar systems  (cf.~\cite{Carlos19,AlmeidaIEEE20,Carlos2018,hay95,AlmeidaIET20} for more discussion on this). 
In~\cite{hay95}, the authors consider the most general case of the Fox's $H$-function for several variables, defined as 
\begin{equation}
\label{eq:FoxDefinition1}
\mathbf{H} \left[ \textbf{x};\delta; \textbf{D} ;\beta; \textbf{B} ; \mathcal{L}_\textbf{s} \right] \triangleq \left(\frac{1}{2 \pi i} \right)^L \oint_{\mathcal{L}_{\textbf{s}}} \Theta \left(  \textbf{s} \right)\textbf{x}^{-\textbf{s}} \text{d} \textbf{s},
\end{equation}
\normalsize
in which $i=\sqrt{-1}$ is the imaginary unit, $\textbf{s}\triangleq\left[ s_1, \cdots,s_L  \right]$, $ \textbf{x}\triangleq \left[ x_1,\cdots,x_L   \right]$, $\beta\triangleq \left[ \beta_1,\cdots, \beta_L \right]$,  and $\delta \triangleq \left[  \delta_1, \cdots, \delta_L \right]$ denote vectors of complex numbers, and $\textbf{B}\triangleq \left(b_{j,l}  \right)_{q \times L}$ and $\textbf{D}\triangleq\left( d_{j,l} \right)_{p \times L}$ are matrices of real numbers. Also, $\textbf{x}^{-\textbf{s}}\triangleq \prod_{l=1}^{L} x_l^{- s_l}$, $\text{d} \textbf{s}\triangleq\prod_{l=1}^{L} \text{d} s_l$, $\mathcal{L}_{\textbf{s}}\triangleq\mathcal{L}_{\textbf{s},1} \times \cdots \times \mathcal{L}_{\textbf{s},L}$, $\mathcal{L}_{\textbf{s},l}$ is an appropriate contour on the complex plane $s_l$, and
\begin{equation}
\label{eq:FoxDefinition2}
 \Theta \left(  \textbf{s} \right) \triangleq \frac{\prod _{j=1}^p \Gamma \left(\delta _j+\sum_{l=1}^L d_{j,l} s_l\right)}{\prod _{j=1}^q \Gamma \left(\beta _j+\sum_{l=1}^L b_{j,l} s_l\right)},
\end{equation}
\normalsize
in which $\Gamma (\cdot)$ is the gamma function~\cite[Eq. (6.1.1)]{abramowitz72}.

\section{System Model}
\label{sec:System Model}
In this section, we describe the standard  system model for a non-coherent detector.

Taking into account the target echo and background noise, the overall complex received signal $r(t)$ can be written as
\begin{align}
    \label{}
    r(t) =s(t)+w(t),
\end{align}
where $s(t)$ denotes the complex target echo, defined as
\begin{align}
    \label{}
    s(t) =\sum_{n=0}^{N-1} A_n \exp \left(i \theta_n \right) p\left(t - n \text{PRI} \right),
\end{align}
in which $p(t)$ represents the unit energy baseband equivalent of each transmitted pulse, $N$ is the number of pulses used for non-coherent integration, $\text{PRI}$ is the pulse repetition interval, $\theta_n$ is the resulting phase corresponding to the $n$-th pulse, $A_n$ is the $n$-th received envelope accounting for propagation effects as well as for target reflectivity, and $w(t)$ is the additive disturbance component modeled as a zero-mean complex circular white Gaussian process.
\begin{figure*}[t]
\begin{center}
\includegraphics[trim={0cm 0cm 0cm 0cm}, clip, scale=0.5]{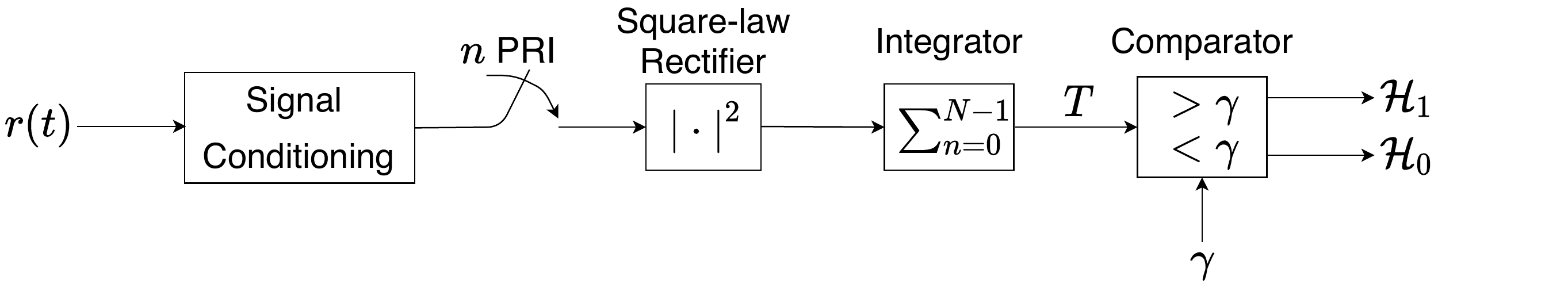}
\caption{Non-coherent detection scheme.}
\label{fig:non-coherent detection scheme} 
\end{center}
\hrulefill
\end{figure*}
In a non-coherent detector, the presence or absent of a target  relies on the following \textit{binary hypothesis test}~\cite{Cui14}:
\begin{subequations}
\begin{align}
    \label{eq:T H1}
    \mathcal{H}_1: \ &T=\sum _{n=0}^{N-1} | A_n \exp \left(i  \theta _n\right) + w_n|^2 \\ \label{eq:T H0}
    \mathcal{H}_0: \ & T=\sum _{n=0}^{N-1} | w_n|^2,
\end{align}
\end{subequations}
where $T$ is the system's test statistics, and $w_n$ is the $n$-th noise sample. The non-coherent detector scheme is depicted in Fig.~\ref{fig:non-coherent detection scheme}.

Radar performance is governed by the PD and PFA.
These probabilities can be computed as the probability that the decision variable $T$, defined respectively as in~\eqref{eq:T H1} and~\eqref{eq:T H0}, falls above the decision threshold, say $\gamma$, i.e.,
\begin{align}
\label{eq:def_pd}
P_{\text{D}} & \triangleq \int_{\gamma}^{\infty } \mathit{f}_T \left(  t| \mathcal{H}_1 \right) \, \text{d} t \\
\label{eq:def_pfa}
P_{\text{FA}} & \triangleq \int_{\gamma }^{\infty } \mathit{f}_T\left(t| \mathcal{H}_0\right) \, \text{d} t.
\end{align}
Consider for the moment that $A_n$ is modeled as a \textit{nonfluctuating} target\footnote{A \textit{nonfluctuating} target (also called Swerling 0 target model) simply means that the target \textit{radar cross section} (RCS) exhibits no random behavior~\cite{richards10}.}, and that $\theta_n$ is modeled as sequence of independent uniformly distributed RVs, i.e., $\theta_n\sim \mathcal{U}\left(0,2 \pi\right)$.
Under these conditions, the PD is given by~\cite{richards10}
\begin{align}
    \label{eq: Q-Def}
    P_{\text{D}} = & Q_{N}\left(\sqrt{2 \zeta},\sqrt{2 \gamma }\right),
\end{align}
where $Q_{(\cdot)}\left(\cdot,\cdot\right)$ is the Marcum's Q-function~\cite{kay98}, and
\begin{align}
    \label{eq: Zeta}
    \zeta=\frac{1}{2 \sigma^2} \sum_{n=0}^{N-1} \xi_n,
\end{align}
with $\xi_n =A_n^2$ being the target power at the $n$-th pulse, and $2 \sigma^2$ being the total noise power accounting for the in-phase and quadrature components. From \eqref{eq: Zeta}, the signal-to-noise ratio (SNR) can be defined as
\begin{align}
    \label{eq: SNR}
    \nonumber \text{SNR} = & \frac{1}{2 \sigma^2} \sum_{n=0}^{N-1}  \mathbb{E} \left[ \xi_n \right] \\
    = & \frac{1}{2 \sigma^2} \sum_{n=0}^{N-1}   \tilde{\Omega }_n^{\frac{1}{\tilde{\alpha }_n}} \Gamma \left(1+\frac{1}{\tilde{\alpha }_n}\right).
\end{align}
On the other hand, the PFA can be calculated as~\cite{Cui14}
\begin{align}
    \label{eq: PFA}
    P_{\text{FA}}=\frac{\Gamma (N,\gamma )}{\Gamma(N)},
\end{align}
in which $\Gamma (\cdot,\cdot)$ is the incomplete gamma function~\cite[Eq. (8.2.1)]{abramowitz72}.
In subsequent sections, we will compute the PD by allowing for Weibull target fluctuations.

\section{Sum Statistics}
\label{sec: Sum Statistics}
In this section, we revisit key results on exact and approximate solutions for the sum of Weibull variates.

First, we define $\eta$ as the sum of $N$ independent RVs $\xi_n$, i.e.,
\begin{align}
    \label{eq: Sum definition}
    \eta = \sum_{n=0}^{N-1} \xi_n.
\end{align}

\subsection{Exact Sum}
\label{sec: Exact Sum of Weibull Variates Variates}
Let $\{\xi_n \}_{n=0}^{N-1}$ be a set of $N$ independent and non-identically
distributed (i.n.i.d) Weibull variates. The PDF of $\xi_n$ given by
\begin{align}
    \label{eq: Weibulls PDFs}
    f_{\xi_n} \left( \xi_n\right) = \frac{\tilde{\alpha}_n \xi_n^{\tilde{\alpha}_n-1}}{\tilde{\Omega}_n}\exp \left(- \frac{\xi_n^{\tilde{\alpha}_n}}{\tilde{\Omega}_n} \right),
\end{align}
where $\tilde{\alpha}_n>0$ is the shape parameter and $\tilde{\Omega}_n =\mathbb{E}\left[\xi_n^{\tilde{\alpha}_n}\right]$ is the scale parameter. In particular, for $\tilde{\alpha}_n=1$ and $\tilde{\alpha}_n=2$, \eqref{eq: Weibulls PDFs} reduces to the Exponential and Rayleigh PDFs, respectively.
Then, the PDF of \eqref{eq: Sum definition} can be written as~\cite{Yilmaz09conf}
\begin{align}
    \label{eq: PDF eta Exact}
    f_{\eta} \left( \eta\right) = \frac{\eta^{N-1}}{\chi^N \Gamma(N)} \sum_{l=0}^{\infty}  \, _1F_1\left(N+l;N;- \frac{\eta}{\chi}\right) a_l, \ \ \ \ \ \eta \geq 0
\end{align}
where $_1F_1\left(\cdot;\cdot;\cdot\right)$ is the Kummer confluent hypergeometric
function~\cite[Eq. (13.1.2)]{Olver10}, and the coefficients $a_l$ and $\chi$ are given, respectively, by
\begin{align}
     \label{eq: al}
    & a_l= \sum_{k_0+\hdots+k_{N-1}=l} \prod_{n=0}^{N-1} \mathcal{V} \left( \xi_n \left| \frac{1}{\chi} \right.\right) \\     \label{eq: Xi}
    & \chi = \frac{2}{N} \sum_{n=0}^{N-1} \tilde{\Omega}_n \\
    & \mathcal{V} \left( \xi_n \left| \frac{1}{\chi} \right.\right) = \sum_{k=0}^{k_n} \frac{(-1)^k \tilde{\Omega}_n^{\frac{k}{\tilde{\alpha}_n}}}{\chi^k k!} \binom{k_n}{k_n-k}  \Gamma \left(\frac{k+\tilde{\alpha}_n}{\tilde{\alpha}_n}\right),
\end{align}
where $\sum_{k_0+\hdots+k_{N-1}=l}$ denotes the summation over all the possible non-negative integers $k_0,\hdots,k_{N-1}$ satisfying the condition $k_0+\hdots+k_{N-1}=l$.
Observe that for a proper calculation, \eqref{eq: PDF eta Exact} requires: \textbf{1)} two infinite sums, in which one of them has to fulfill some impositions; \textbf{2)} $N$ finite sums for each interaction; and \textbf{3)} $N$ products for each interaction.
More importantly, observe that the mathematical complexity  of \eqref{eq: PDF eta Exact} increases as $N$ increases.

For the case of independent and identically distributed (i.i.d) Weibull variates (i.e., $\tilde{\alpha}_n=\tilde{\alpha}_n,\tilde{\Omega}_n=\tilde{\Omega}$), the PDF of $\eta$ is still given by \eqref{eq: PDF eta Approx}, however, the coefficients $a_l$ and $\chi$ are now defined, respectively, as
\begin{align}
    \label{}
    & a_l= \sum_{k_0+\hdots+k_{N-1}=l} \prod_{n=0}^{N-1} \mathcal{V} \left( \xi_n \left| \frac{1}{\chi} \right.\right) \\
    & \chi = 2 \ \tilde{\Omega}\\
    & \mathcal{V} \left( \xi_n \left| \frac{1}{\chi} \right.\right) = \sum_{k=0}^{k_n} \frac{(-1)^k \tilde{\Omega}^{\frac{k}{\tilde{\alpha}}}}{\chi^k k!} \binom{k_n}{k_n-k}  \Gamma \left(\frac{k+\tilde{\alpha}}{\tilde{\alpha}}\right).
\end{align}

\subsection{Approximate Sum}
\label{sec: Sum Approximation}
In~\cite{Filho06}, a simple and accurate approximation for the sum of i.i.d Weibull variates was derived. The authors proposed to approximate the sum in \eqref{eq: Sum definition} by the $\alpha$-$\mu$ envelope, given by~\cite{Yacoub07}
\begin{align}
    \label{eq: PDF eta Approx}
    f_{\eta} \left( \eta\right)= \frac{\alpha\mu^{\mu} \eta^{\alpha \mu -1}}{\Omega^\mu \Gamma (\mu)} \exp \left( - \frac{\mu \eta^{\alpha}}{\Omega} \right),
\end{align}
where $\alpha>0$ is the shape parameter, $\Omega =\mathbb{E}\left[\eta^{\alpha}\right]$ is the scale parameter, and $\mu=\mathbb{E}^2\left[\eta^{\alpha}\right]/\mathbb{V} \left[\eta^{\alpha} \right]>0$ is the inverse normalized variance of $\eta^\alpha$.
This approximation has been anchored in the fact that the $\alpha$-$\mu$ envelope is modeled as the $\alpha$-root of the sum of i.i.d. squared Rayleigh variates, resembling somehow the algebraic structure of the exact Weibull sum, in which the $n$-th summand can be written as the $\tilde{\alpha}_n$-root of a squared
Rayleigh variate~\cite{Yacoub02}.

In order to render \eqref{eq: PDF eta Approx} a good approximation, the moment-based estimators~\cite{papoulis02} is applied for $\Omega$, $\alpha$ and $\mu$, i.e.,
\begin{align}
    \label{eq: Match 1}
    \frac{\mathbb{E}^2\left[\eta \right]}{\mathbb{E} \left[\eta^2 \right]- \mathbb{E}^2 \left[\eta\right]} =& \frac{\Gamma^2(\mu+\frac{1}{\alpha})}{\Gamma(\mu) \Gamma(\mu +\frac{2}{\alpha})-\Gamma^2(\mu +\frac{1}{\alpha})}\\ \label{eq: Match 2}
   \frac{\mathbb{E}^2\left[\eta^2 \right]}{\mathbb{E} \left[\eta^4 \right]- \mathbb{E}^2 \left[\eta^2 \right]} =&  \frac{\Gamma^2(\mu+\frac{2}{\alpha})}{\Gamma(\mu) \Gamma(\mu +\frac{4}{\alpha})-\Gamma^2(\mu +\frac{2}{\alpha})}\\ \label{eq: Match 3}
   \Omega =& \left[ \frac{\mu^{1/\alpha} \Gamma(\mu) \mathbb{E} \left[\eta \right]}{\Gamma(\mu + \frac{1}{\alpha})}\right]^\alpha.
\end{align}
\normalsize
The exact moments $\mathbb{E} \left[ \eta\right]$, $\mathbb{E} \left[ \eta^2\right]$ and $\mathbb{E} \left[ \eta^4 \right]$ can be obtained through the multinomial expansion as~\cite{Kreyszig10}
\begin{align}
    \label{eq: Moments eta}
    \nonumber \mathbb{E} \left[ \eta^p \right]=& \sum_{p_1=0}^{p}\sum_{p_2=0}^{p_1} \cdots\sum_{p_{N-2}=0}^{p_{N-3}} \binom{p}{p_1}\binom{p_1}{p_2} \cdots \binom{p_{N-3}}{p_{N-2}} \\
    & \times \mathbb{E} \left[ \xi_{0}^{p-p_1} \right] \mathbb{E} \left[ \xi_{1}^{p_1-p_2} \right] \cdots \mathbb{E} \left[ \xi_{N-1}^{p_{N-2}} \right],
\end{align}
where $p$ is a positive integer and the required Weibull moments are given by
\begin{align}
    \label{eq: Moments xi}
    \mathbb{E} \left[ \xi_{n}^{p} \right] = \tilde{\Omega}_n^{\frac{p}{\tilde{\alpha}_n}} \Gamma \left(1+\frac{p}{\tilde{\alpha}_n} \right).
\end{align}

\section{Detection Performance}
\label{sec: Detection Performance}
In this section, we derive the PD by modeling $\xi_n$ as a set of i.i.d. Weibull RVs.

To do so, we first derive the PDF of $\zeta$. This can be easily obtained by performing a transformation of variables in \eqref{eq: PDF eta Approx}, resulting in
\begin{align}
    \label{eq: PDF zeta Approx}
    \mathit{f}_\zeta \left(\zeta \right) = \frac{\alpha  \mu ^{\mu }  \left(2 \zeta  \sigma ^2\right)^{\alpha  \mu } \exp \left(-\frac{\mu  \left(2 \zeta  \sigma ^2\right)^{\alpha }}{\Omega }\right)}{ \zeta \Omega^{\mu }   \Gamma (\mu )}.
\end{align}
Now, by using \eqref{eq: Q-Def} and \eqref{eq: PDF eta Approx}, the  PD can be defined as
\footnote{The sub-index $\mathcal{W}$ in \eqref{eq:PD_def} refers to the use of the Weibull fluctuating target model.}
\begin{align}
    \label{eq:PD_def}
    P_{\text{D}_{\mathcal{W}}} \triangleq \int_0^{\infty } Q_{N}\left(\sqrt{2 \zeta},\sqrt{2 \gamma }\right) \mathit{f}_\zeta \left(\zeta \right) \, \text{d}\zeta.
\end{align}
In order to solve \eqref{eq:PD_def}, we start by using the Marcum's Q-function definition~\cite[Eq. (15.2)]{richards10}:

\begin{align}
    \label{eq: Marcum}
    \nonumber Q_{N}\left(\sqrt{2 \zeta},\sqrt{2 \gamma }\right) =& \int_{\sqrt{2 \gamma }}^{\infty } x \exp \left(-\frac{1}{2} \left(x^2+2 \zeta\right)\right) \\
    & \times \left(\frac{x}{\sqrt{2 \zeta}}\right)^{N-1} I_{N-1}\left(\sqrt{2 \zeta} x\right) \, \text{d}x,
\end{align}
where $I_{(\cdot)}(\cdot)$ is modified Bessel function of the first kind~\cite[Eq. (03.02.02.0001.01)]{Mathematica}.

Replacing \eqref{eq: PDF zeta Approx} and \eqref{eq: Marcum} in \eqref{eq:PD_def}, yields
\begin{align}
    \label{}
    \nonumber P_{\text{D}_{\mathcal{W}}} = & \frac{2 \alpha  \mu ^{\mu } \sigma ^2 \left(2 \sigma ^2\right)^{\alpha  \mu -1}}{\Omega ^{\mu } \Gamma (\mu )}\int _0^{\infty }\int _{\sqrt{2 \gamma }}^{\infty }x  \zeta ^{\alpha  \mu -1} \left(\frac{x}{\sqrt{2 \zeta }}\right)^{N-1} \\
    \nonumber & \times \exp \left(-\frac{1}{2} \left(2 \zeta +x^2\right)\right) I_{N-1}\left(\sqrt{2 \zeta } x\right) \\
    & \times \exp \left(-\frac{\mu  \left(2 \zeta  \sigma ^2\right)^{\alpha }}{\Omega }\right) \text{d}x \ \text{d}\zeta.
\end{align}
Since $\int_0^{\infty } |Q_{N}\left(\sqrt{2 \zeta},\sqrt{2 \gamma }\right) \mathit{f}_\zeta \left(\zeta \right)| \, \text{d}\zeta < \infty$, we can invoke the Fubini's theorem~\cite{fubibi07} so as to interchange the order of integration, i.e.,
\begin{align}
    \label{eq: PD 1}
    \nonumber P_{\text{D}_{\mathcal{W}}} =& \frac{\alpha  \mu ^{\mu } \left(2 \sigma ^2\right)^{\alpha  \mu }}{\Omega ^{\mu } \Gamma (\mu )}\int _{\sqrt{2 \gamma }}^{\infty }x \exp \left(-\frac{x^2}{2}\right) \left(\frac{x}{\sqrt{2  }}\right)^{N-1} \\
    \nonumber & \times  \int _0^{\infty }\zeta ^{\alpha  \mu-1 - (N-1)/2} \exp (-\zeta )  I_{N-1}\left(\sqrt{2 \zeta } x\right) \\
    & \times  \exp \left(-\frac{\mu  \left(2 \zeta  \sigma ^2\right)^{\alpha }}{\Omega }\right) \text{d}\zeta \ \text{d} x.
\end{align} 
Now, by making use of \cite[Eq. (03.02.26.0007.01)]{Mathematica} and \cite[Eq. (01.03.26.0004.01)]{Mathematica}, we can rewrite \eqref{eq: PD 1} as
\begin{align}
    \label{eq:}
    \nonumber P_{\text{D}_{\mathcal{W}}} =& \frac{\alpha  \mu ^{\mu } \left(2 \sigma ^2\right)^{\alpha  \mu }}{\Omega ^{\mu } \Gamma (\mu )} \int _{\sqrt{2 \gamma }}^{\infty }x \exp \left(-\frac{x^2}{2}\right) \left(\frac{x}{\sqrt{2  }}\right)^{N-1} \\
    \nonumber & \times  \int _0^{\infty }\zeta ^{\alpha  \mu-1 - (N-1)/2} \exp (-\zeta ) \ i^{1-N}   \\
    \nonumber & \times G_{0,2}^{1,0} \left[ \begin {array} {c} - \\ \frac{N-1}{2},\frac{1-N}{2} \\\end {array} \left| -\frac{\zeta  x^2}{2}  \right. \right] \\
    & \times G_{1,0}^{0,1} \left[ \begin {array} {c} - \\0 \\\end {array} \left| \frac{\mu  \left(2 \zeta  \sigma ^2\right)^{\alpha }}{\Omega } \right. \right] \text{d}\zeta \ \text{d} x,
\end{align}
where $G_{m,n}^{p,q} \left[ \cdot \right]$ \normalsize is the Meijer's G-function~\cite[Eq. (16.17.1)]{Olver10}.

Then, using the contour integral representation of the Meijer's G-function~\cite[Eq. (07.34.02.0001.01)]{Mathematica}, along with some mathematical manipulations, we obtain
\begin{align}
    \label{}
    \nonumber P_{\text{D}_{\mathcal{W}}} =& \frac{\alpha  \mu ^{\mu } \left(2 \sigma ^2\right)^{\alpha  \mu }}{\Omega ^{\mu } \Gamma (\mu )} \int _{\sqrt{2 \gamma }}^{\infty }x \exp \left(-\frac{x^2}{2}\right) \left(\frac{x}{\sqrt{2}}\right)^{N-1}  \\
    \nonumber & \times  \left(\frac{1}{2 \pi  i}\right)^2 \oint_{\ddot{\mathcal{L}}_{\textbf{s},1}} \oint_{\ddot{\mathcal{L}}_{\textbf{s},2}} \frac{i^{3 N+1} \Gamma \left(s_1\right) \Gamma \left(\frac{N-1}{2}+s_2\right)}{\Gamma \left(\frac{N-1}{2}-s_2+1\right)} \\
    \nonumber & \times  \left(\frac{\mu  \left(2 \sigma ^2\right)^{\alpha }}{\Omega }\right)^{-s_1} \left(-\frac{x^2}{2}\right)^{-s_2} \int _0^{\infty }\exp (-\zeta ) \\
    & \times \zeta ^{\alpha  \mu -\alpha  s_1-s_2-1-(N-1)/2} \text{d}\zeta \ \text{d}s_1 \ \text{d}s_2 \ \text{d}x,
\end{align}
where $\ddot{\mathcal{L}}_{\textbf{s},1}$ and $\ddot{\mathcal{L}}_{\textbf{s},2}$ are suitable contours in the complex plane. 

Now, developing the inner integral and reordering the order of integration, yields
\begin{align}
    \label{eq: L1 L2 ast}
    \nonumber P_{\text{D}_{\mathcal{W}}} &=  \frac{\alpha  \mu ^{\mu } \sqrt{2}^{1-N} \left(2 \sigma ^2\right)^{\alpha  \mu }}{\Omega ^{\mu } \Gamma (\mu )} \left(\frac{1}{2 \pi  i}\right)^2 \oint_{\dot{\mathcal{L}}_{\textbf{s},1}} \oint_{\dot{\mathcal{L}}_{\textbf{s},2}} i^{3 N+1} \\
    \nonumber & \times \frac{\Gamma \left(s_1\right) \Gamma \left(\frac{N-1}{2}+s_2\right) \Gamma \left(-\frac{N}{2}+\alpha  \mu -\alpha  s_1-s_2+\frac{1}{2}\right)}{\Gamma \left(\frac{N-1}{2}-s_2+1\right)} \\
    \nonumber & \times  \left(\frac{\mu  \left(2 \sigma ^2\right)^{\alpha }}{\Omega }\right)^{-s_1} \left(-\frac{1}{2}\right)^{-s_2}   \\
    & \times \int _{\sqrt{2 \gamma }}^{\infty } x^{N-2 s_2} \exp \left(-\frac{x^2}{2}\right) \text{d}x \ \text{d}s_1 \ \text{d}s_2.
\end{align}
in which $\dot{\mathcal{L}}_{\textbf{s},1}$ and $\dot{\mathcal{L}}_{\textbf{s},2}$ are two new suitable contours.
They appear since the last integration deformed the integration paths of $\ddot{\mathcal{L}}_{\textbf{s},1}$ and $\ddot{\mathcal{L}}_{\textbf{s},2}$. 
\begin{table*}[t]
\centering
\caption{Arguments for the Fox's $H$-functions.}
\begin{tabular}{c c c c c c}
\hline 
\hline 
\begin{tabular}[c]{@{}c@{}}\textcolor{white}{.} \\ \textcolor{white}{.} \end{tabular} $\textbf{x}^{\dagger}$  & \hspace{-0.2cm} $\delta^{\dagger}$ & \hspace{-0.2cm} $\textbf{D}^{\dagger}$ & \hspace{-0.2cm} $\beta^{\dagger}$ & \hspace{-0.2cm} $\textbf{B}^{\dagger}$ & \hspace{-0.2cm} $\mathcal{L}_{\textbf{s}}^{\dagger}$\\ \hline $\left[\frac{\mu  \left(2 \sigma ^2\right)^{\alpha }}{\Omega },-1\right]$ & \hspace{-0.2cm} $\left[0,\frac{N-1}{2},\alpha  \mu -\frac{N}{2}+\frac{1}{2},\frac{N}{2}+\frac{1}{2}\right]$ &  \hspace{-0.2cm} $\left(
\begin{array}{cc}
 1 & 0 \\
 0 & 1 \\
 -\alpha  & -1 \\
 0 & -1 \\
\end{array}
\right)$ & \hspace{-0.2cm} $\left[\frac{N-1}{2}+1\right]$ &  $\left(
\begin{array}{cc}
 0 & -1 \\
\end{array}
\right)$ & \hspace{-0.2cm} $\mathcal{L}_{\textbf{s},1}^{\dagger} \times \mathcal{L}_{\textbf{s},2}^{\dagger}$ \\  \hline \hline
\begin{tabular}[c]{@{}c@{}}\textcolor{white}{.} \\ \textcolor{white}{.} \end{tabular} $\textbf{x}^{\ddagger}$  & \hspace{-0.2cm} $\delta^{\ddagger}$ & \hspace{-0.2cm}  $\textbf{D}^{\ddagger}$ & \hspace{-0.2cm} $\beta^{\ddagger}$ &  $\textbf{B}^{\ddagger}$ & \hspace{-0.2cm} $\mathcal{L}_{\textbf{s}}^{\ddagger}$ \\ \hline
$\left[\frac{\mu  \left(2 \sigma ^2\right)^{\alpha }}{\Omega },-1,\gamma\right]$ & \hspace{-0.2cm} $\left[0,\frac{N-1}{2},\alpha  \mu -\frac{N}{2}+\frac{1}{2},\frac{N}{2}+\frac{1}{2},0\right]$ & \hspace{-0.2cm} $\left(
\begin{array}{ccc}
 1 & 0 & 0 \\
 0 & 1 & 0 \\
 -\alpha  & -1 & 0 \\
 0 & -1 & 1 \\
 0 & 0 & -1 \\
\end{array}
\right)$ & \hspace{-0.2cm} $\left[\frac{N-1}{2}+1,1\right]$ & \hspace{-0.2cm} $\left(
\begin{array}{cc}
 0 & 0 \\
 -1 & 0 \\
 0 & -1 \\
\end{array}
\right)^T$ & \hspace{-0.2cm} $\mathcal{L}_{\textbf{s},1}^{\ddagger} \times \mathcal{L}_{\textbf{s},2}^{\ddagger} \times \mathcal{L}_{\textbf{s},3}^{\ddagger}$ \\ \hline \hline
\end{tabular}
\label{tab: Table1}
\end{table*}
\begin{figure}[t]
\begin{center}
\includegraphics[trim={0cm 0cm 5cm 0cm},clip,scale=0.38]{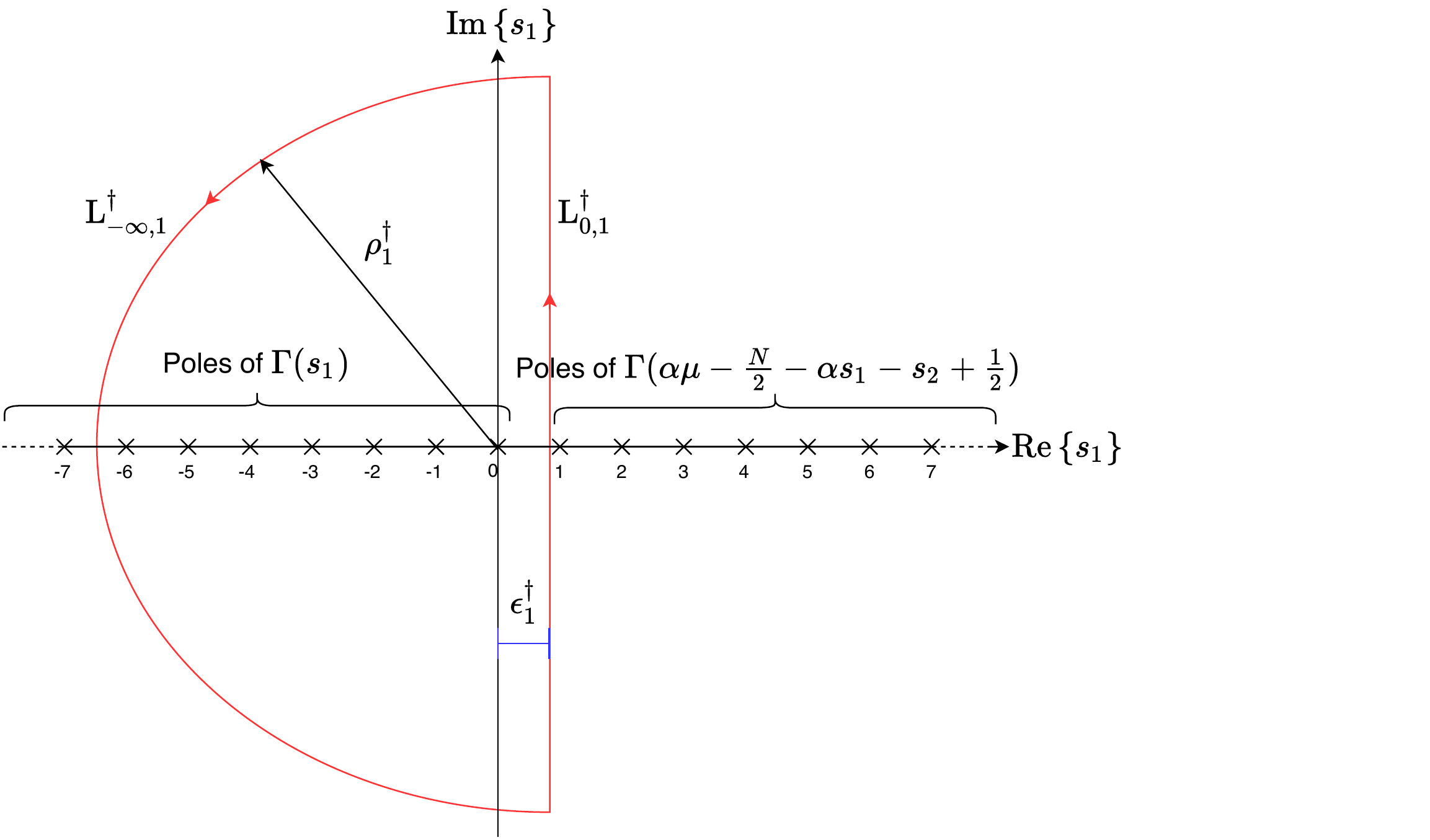}
\caption{Integration path for $\mathcal{L}_{\textbf{s},1}^{\dagger}$.}
\label{fig: L1 dag}
\end{center} 
\end{figure}
\begin{figure}[t]
\begin{center}
\includegraphics[trim={0cm 0cm 5cm 0cm},clip,scale=0.38]{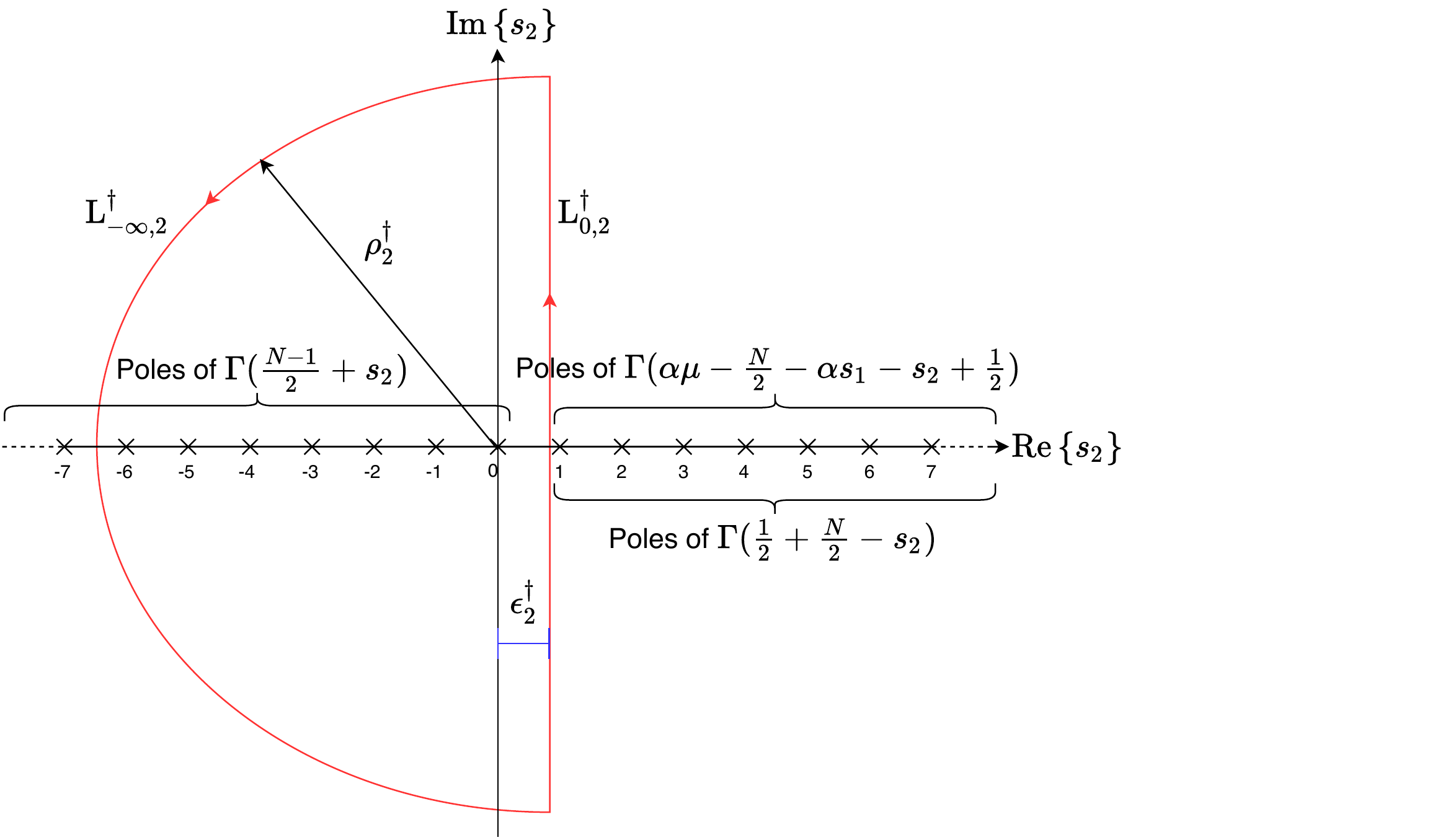}
\caption{Integration path for $\mathcal{L}_{\textbf{s},2}^{\dagger}$.}
\label{fig: L2 dag}
\end{center} 
\end{figure}
\begin{figure}[t]
\begin{center}
\includegraphics[trim={0cm 0cm 5cm 0cm},clip,scale=0.38]{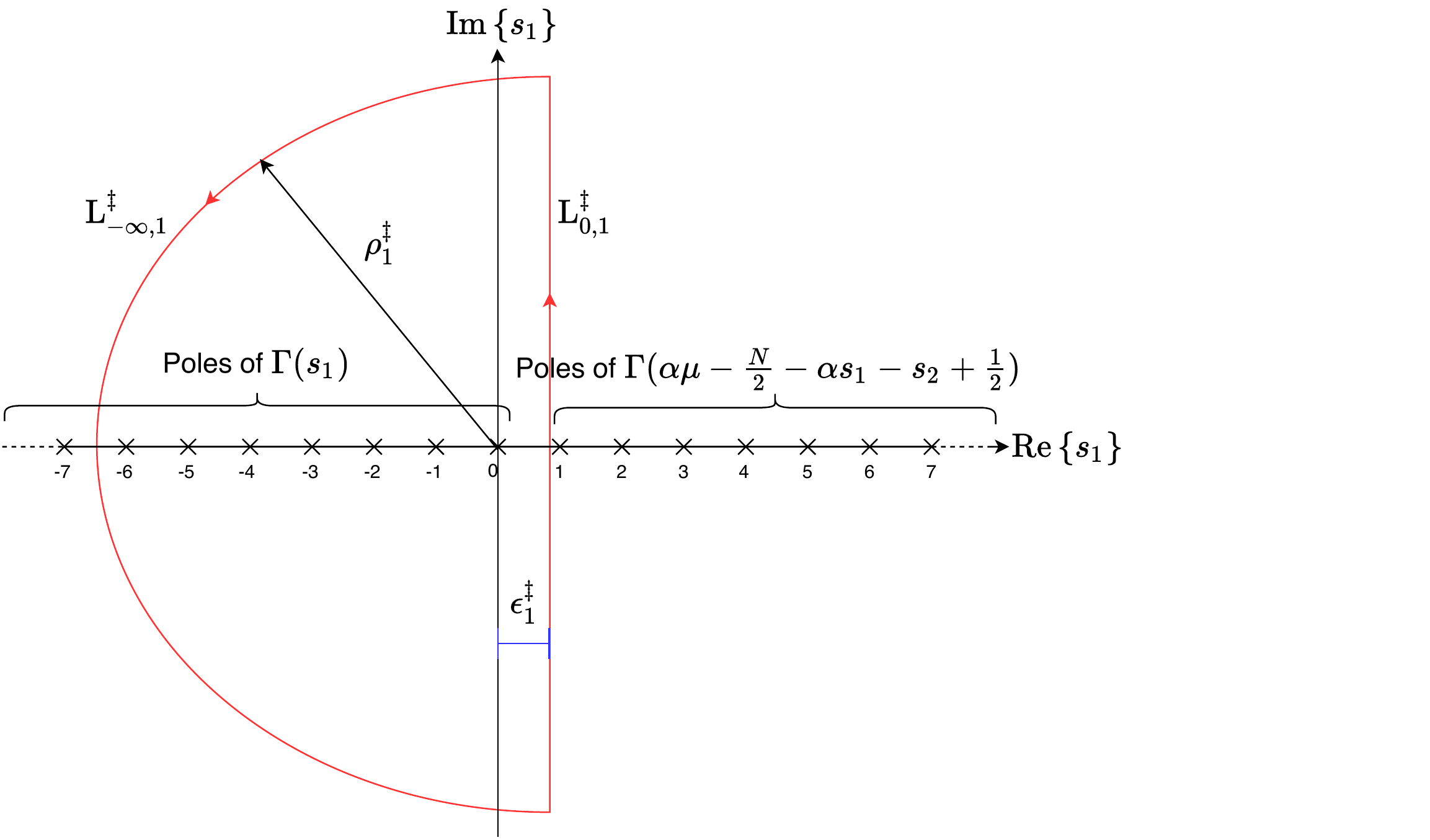}
\caption{Integration path for $\mathcal{L}_{\textbf{s},1}^{\ddagger}$.}
\label{fig: L1 ast}
\end{center} 
\end{figure}
\begin{figure}[t]
\begin{center}
\includegraphics[trim={0cm 0cm 5cm 0cm},clip,scale=0.38]{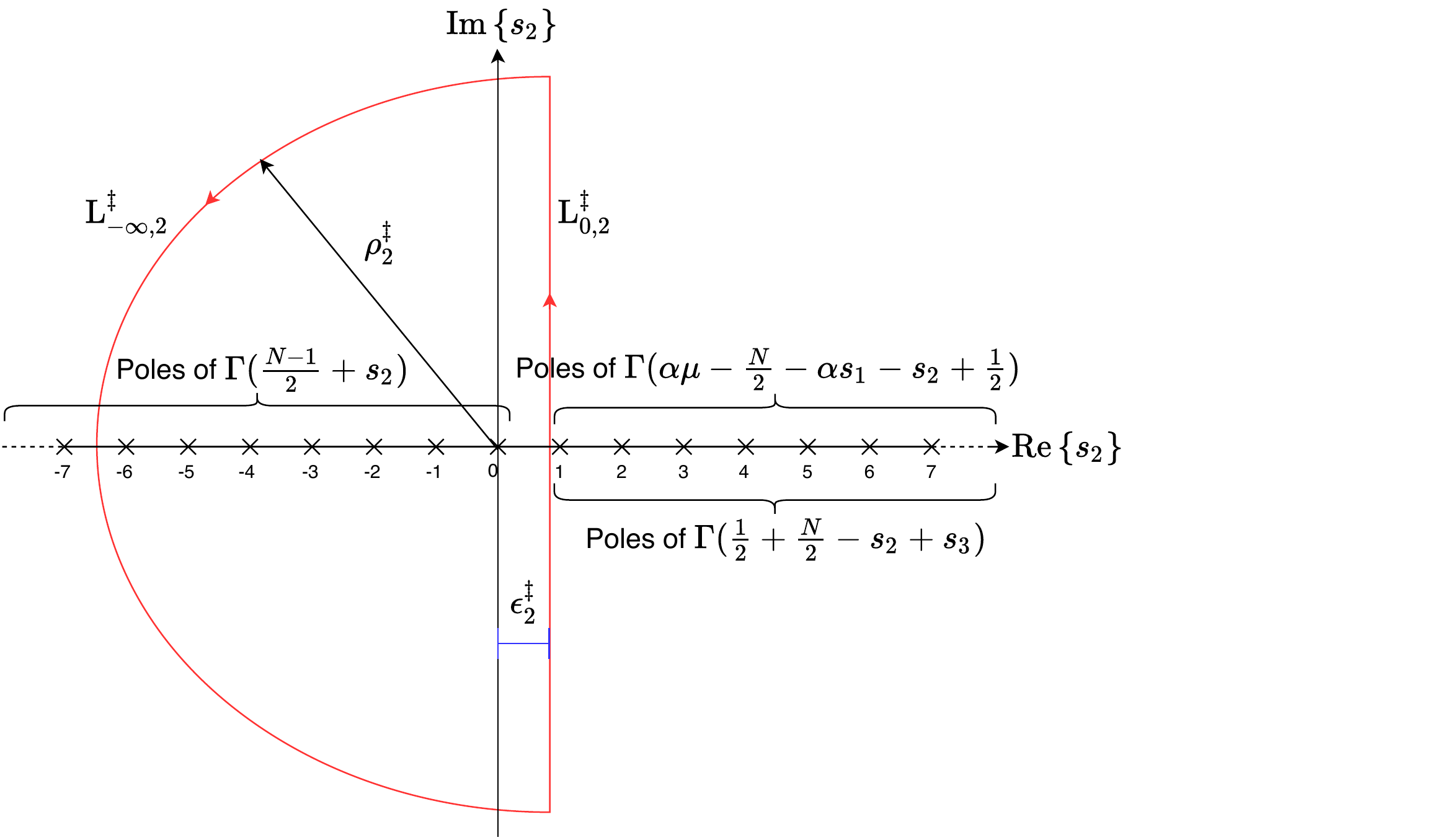}
\caption{Integration path for $\mathcal{L}_{\textbf{s},2}^{\ddagger}$.}
\label{fig: L2 ast}
\end{center} 
\end{figure}
\begin{figure}[t]
\begin{center}
\includegraphics[trim={0cm 0cm 4cm 0cm},clip,scale=0.38]{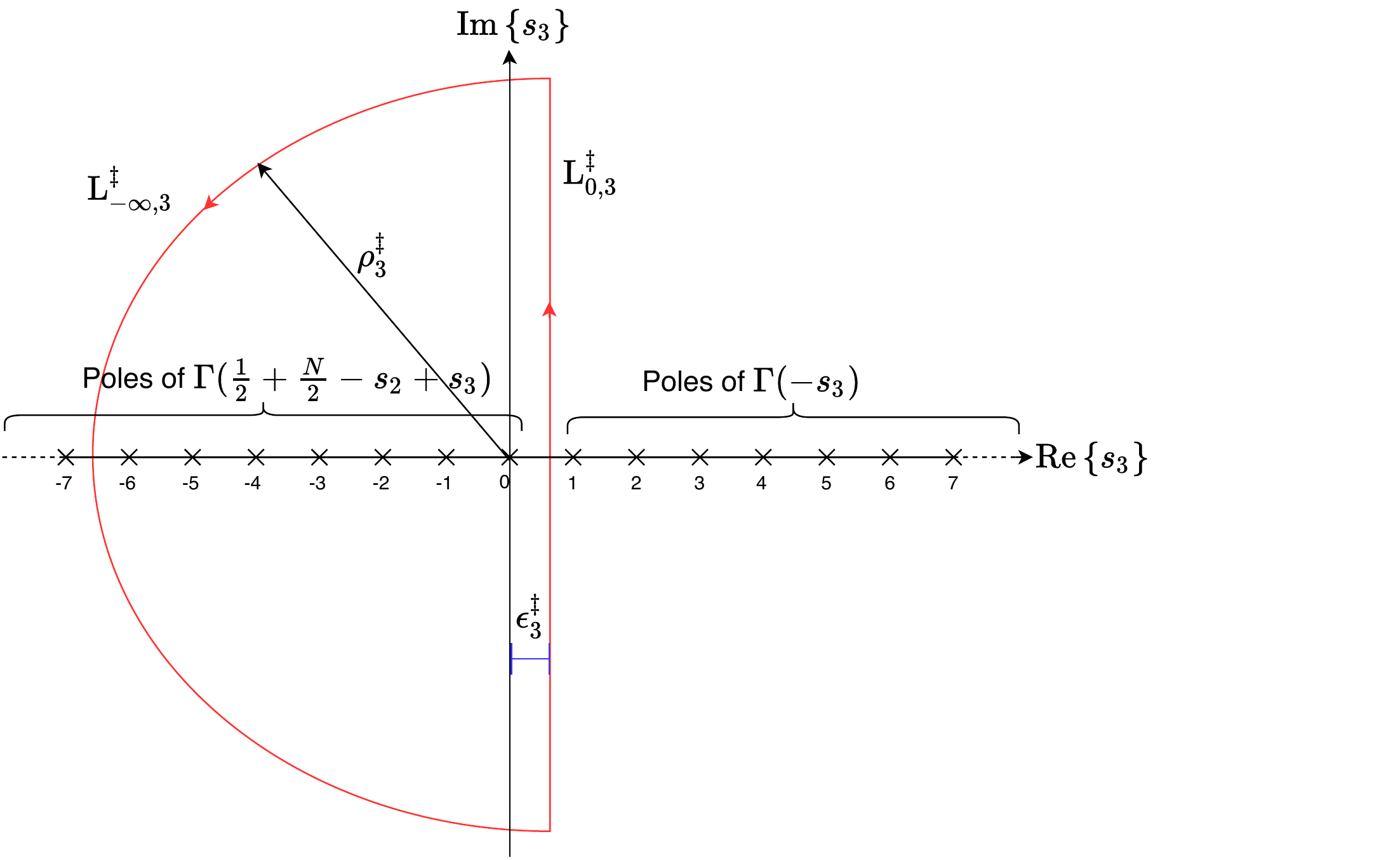}
\caption{Integration path for $\mathcal{L}_{\textbf{s},3}^{\ddagger}$.}
\label{fig: L3 ast}
\end{center} 
\end{figure}

Finally, evaluating the remaining integral with the aid of \cite[Eq. (06.06.07.0002.01)]{Mathematica}, and followed by lengthy mathematical manipulations, we obtain a closed-form solution for \eqref{eq:PD_def} given by
\begin{align}
    \label{eq: Final FOX}
    \nonumber P_{\text{D}_{\mathcal{W}}} =& \Phi \left( \textbf{H} \left[ \textbf{x}^{\dagger}; \delta^{\dagger} ; \textbf{D}^{\dagger} ;\beta^{\dagger}; \textbf{B}^{\dagger} ; \mathcal{L}_{\textbf{s}}^{\dagger} \right] \right. \\
    & - \left. \textbf{H}\left[ \textbf{x}^{\ddagger} ; \delta^{\ddagger} ; \textbf{D}^{\ddagger} ; \beta^{\ddagger} ; \textbf{B}^{\ddagger} ; \mathcal{L}_{\textbf{s}}^{\ddagger} \right]  \right),
\end{align}
where $\Phi =\left(\alpha  \mu ^{\mu } i^{1-N} \left(2 \sigma ^2\right)^{\alpha  \mu } \right)/\Omega ^{\mu } \Gamma (\mu )$ and the remaining arguments of the Fox's $H$-function are given in Table~\ref{tab: Table1}.  
In addition, the integration paths for the complex contours defined in \eqref{eq: Final FOX} are listed below:
\begin{itemize}
    \item $\mathcal{L}_{\textbf{s},1}^\dagger$ is a semicircle formed by the segments $\text{L}_{0,1}^\dagger$ and $\text{L}_{-\infty,1}^\dagger$, as shown in Fig.~\ref{fig: L1 dag}, where $\rho_1^\dagger$ is the radius of the semicircle and $\epsilon_1^\dagger$ is a real that must be chosen so that all the poles of $\Gamma(s_1)$ are separated from those of $\Gamma(\alpha \mu -\frac{N}{2}-\alpha s_1-s_2+\frac{1}{2})$.
    \item $\mathcal{L}_{\textbf{s},2}^\dagger$ is a semicircle formed by the segments $\text{L}_{0,2}^\dagger$ and $\text{L}_{-\infty,2}^\dagger$, as shown in Fig.~\ref{fig: L2 dag}, where $\rho_2^\dagger$ is the radius of the semicircle and $\epsilon_2^\dagger$ is a real that must be chosen so that all the poles of $\Gamma(\frac{N-1}{2}+s_2)$ are separated from those of $\Gamma(\alpha \mu -\frac{N}{2}-\alpha s_1-s_2+\frac{1}{2})$ and $\Gamma(\frac{1}{2}+\frac{N}{2}-s_2)$.
    \item $\mathcal{L}_{\textbf{s},1}^\ddagger$ is a semicircle formed by the segments $\text{L}_{0,1}^\ddagger$ and $\text{L}_{-\infty,1}^\ddagger$, as shown in Fig.~\ref{fig: L1 ast}, where $\rho_1^\ddagger$ is the radius of the semicircle and $\epsilon_1^\ddagger$ is a real that must be chosen so that all the poles of $\Gamma(s_1)$ are separated from those of $\Gamma(\alpha \mu -\frac{N}{2}-\alpha s_1-s_2+\frac{1}{2})$.
    \item $\mathcal{L}_{\textbf{s},2}^\ddagger$ is a semicircle formed by the segments $\text{L}_{0,2}^\ddagger$ and $\text{L}_{-\infty,2}^\ddagger$, as shown in Fig.~\ref{fig: L2 ast}, where $\rho_2^\ddagger$ is the radius of the semicircle and $\epsilon_2^\ddagger$ is a real that must be chosen so that all the poles of $\Gamma(\frac{N-1}{2}+s_2)$ are separated from those of $\Gamma(\alpha \mu -\frac{N}{2}-\alpha s_1-s_2+\frac{1}{2})$ and $\Gamma(\frac{1}{2}+\frac{N}{2}-s_2)$ and $\Gamma (\frac{1}{2}+\frac{N}{2}-s_2+s_3)$.
    \item $\mathcal{L}_{\textbf{s},3}^\ddagger$ is a semicircle formed by the segments $\text{L}_{0,3}^\ddagger$ and $\text{L}_{-\infty,3}^\ddagger$, as shown in Fig.~\ref{fig: L3 ast}, where $\rho_3^\ddagger$ is the radius of the semicircle and $\epsilon_3^\ddagger$ is a real that must be chosen so that all the poles of $\Gamma (\frac{1}{2}+\frac{N}{2}-s_2+s_3)$ are separated from those of $\Gamma(-s_3)$.
\end{itemize}
A general implementation for the multivariate Fox's $H$-function is not yet available in mathematical packages such as MATHEMATICA, MATLAB, or MAPLE. 
Some works have been done to alleviate this problem~\cite{alhennawi16,Garcia19,yilmaz09}.
Specifically in~\cite{alhennawi16}, the Fox's $H$-function was implemented from one up to four variables.
In this work, we provide an accurate and portable implementation in MATHEMATICA for the trivariate Fox's $H$-function needed in \eqref{eq: Final FOX}.
This routine can be found in Appendix~\ref{sec: Mathematica Implementation}.
Moreover, an equivalent series  representation for~\eqref{eq: Final FOX} is also provided to ease the computation of our results. 
This series representation is presented in the subsequent subsection.

\section{Alternative Series Representation}
\label{sec: Series Representation}
In this section, we derive a series representation for~\eqref{eq: Final FOX} by means of a thorough calculus of residues.

In order to apply the residue theorem~\cite{Kreyszig10}, all the poles must lie inside the corresponding semicircles. Hence, the radius of each semicircle must tend to infinity.
It can be shown that any complex integration along the paths $\text{L}_{-\infty,1}^\dagger$, $\text{L}_{-\infty,2}^\dagger$, $\text{L}_{-\infty,1}^\ddagger$, $\text{L}_{-\infty,2}^\ddagger$, and $\text{L}_{-\infty,3}^\ddagger$ approaches zero as $\rho_1^\dagger$, $\rho_2^\dagger$, $\rho_1^\ddagger$, $\rho_2^\ddagger$, and $\rho_3^\ddagger$ go to infinity, respectively.
Therefore, the final integration paths will only include a straight lines $\text{L}_{0,1}^\dagger$, $\text{L}_{0,2}^\dagger$, $\text{L}_{0,1}^\ddagger$, $\text{L}_{0,2}^\ddagger$, and $\text{L}_{0,3}^\ddagger$, each of them starting at $- i \infty$ and ending at $i \infty$.

Now, we can rewrite~\eqref{eq: Final FOX} through the sum of residues~\cite{Kreyszig10} as in \eqref{eq: Residue Definition}, shown at the top of the next page, where $\text{Res} \left[ \mathcal{G} \left(a_1,a_2,\hdots,a_p\right);\left\{b_1;b_2;\hdots;b_p\right\}\right]$  denotes the residue of an arbitrary function, say $\mathcal{G} \left(a_1,a_2,\hdots,a_p\right)$, evaluated at the poles  $a_1=b_1$, $a_2=b_2$, $\hdots$ , $a_p=b_p$.
\begin{figure*}[ht]
\begin{flushleft}
\begin{align}
    \label{eq: Residue Definition}
    P_{\text{D}_{\mathcal{W}}} = &\Phi \left[ \sum _{k,l=0}^{\infty }  \text{Res}\left[ \Xi_1 \left(s_1,s_2 \right);\left\{-k;-l-\frac{N}{2}+\frac{1}{2}\right\}\right]  -  \sum _{k,l,m=0}^{\infty }  \text{Res}\left[ \Xi_2 \left(s_1,s_2,s_3 \right);\left\{-k;-l-\frac{N}{2}+\frac{1}{2};-l-m-N\right\}\right] \right]
\end{align}
\end{flushleft}
\hrulefill
\end{figure*} 

In our case, the functions $ \Xi_1$ and $\Xi_2$ in \eqref{eq: Residue Definition} denote the integration kernels of \eqref{eq: Final FOX}, defined , respectively, as
\small
\begin{align}
    \label{}
    \nonumber \Xi_1=&\frac{\Gamma \left(s_1\right) \Gamma \left(\frac{N-1}{2}+s_2\right) \Gamma \left(\alpha  \mu -\frac{N}{2}-\alpha  s_1-s_2+\frac{1}{2}\right)  }{\Gamma \left(\frac{N-1}{2}-s_2+1\right)}\\
    & \times \Gamma \left(\frac{N}{2}-s_2+\frac{1}{2}\right) \left(\frac{\mu  \left(2 \sigma ^2\right)^{\alpha }}{\Omega }\right)^{-s_1} (-1)^{-s_2} \\
    \nonumber \Xi_2=& \frac{\Gamma \left(s_1\right) \Gamma \left(\frac{N-1}{2}+s_2\right) \Gamma \left(-\frac{N}{2}+\alpha  \mu -\alpha  s_1-s_2+\frac{1}{2}\right) \Gamma \left(-s_3\right)}{\Gamma \left(1-s_3\right) \Gamma \left(\frac{N-1}{2}-s_2+1\right)}\\
    & \times  \Gamma \left(\frac{N}{2}-s_2+s_3+\frac{1}{2}\right) \left(\frac{\mu  \left(2 \sigma ^2\right)^{\alpha }}{\Omega }\right)^{-s_1} (-1)^{-s_2} \ \gamma ^{-s_3}.
\end{align}
\normalsize
Applying the residue operation in \eqref{eq: Residue Definition}, we obtain
\begin{align}
    \label{eq: PD I1 I2}
    P_{\text{D}_{\mathcal{W}}} &= \Phi \left[ \mathcal{I}_1-\mathcal{I}_2 \right],
\end{align}
where $\mathcal{I}_1$ and $\mathcal{I}_2$ are summations defined, respectively, by
\small
\begin{align}
    \label{eq: I1}
    \mathcal{I}_1=& \sum _{k,l=0}^{\infty} \frac{i^{N+1} (-1)^{k+1} \Gamma (l+k \alpha +\alpha  \mu )  \left(\frac{\mu  \left(2 \sigma ^2\right)^{\alpha }}{\Omega }\right)^k}{k! \  l!} \\  \label{eq: I2}
    \nonumber \mathcal{I}_2=& \sum _{k,l,m=0}^{\infty } \frac{i^{N+1} (-1)^{k+m+1} \gamma ^{l+m+N} \Gamma (l+m+N) }{k! \ l! \  m! \ \Gamma (l+N) \Gamma (l+m+N+1)}  \\
    & \ \ \  \times \Gamma (l+k \alpha +\alpha  \mu ) \left(\frac{\mu  \left(2 \sigma ^2\right)^{\alpha }}{\Omega }\right)^k.
\end{align}
\normalsize
For convenience, we start by solving $\mathcal{I}_2$. 
Using \cite[Eq. (5.2.8.1)]{prudnikov92} and \cite[Eq. (5.5.1)]{Olver10},  followed by lengthy mathematical manipulations, we can express \eqref{eq: I2} as
\small
\begin{align}
    \label{eq: I2 split}
    \nonumber \mathcal{I}_2 &=  \sum _{k,l=0}^{\infty} \frac{i^{N+1} (-1)^{k+2 l+1} \Gamma (l+k \alpha +\alpha  \mu )  \left(\frac{\mu  \left(2 \sigma ^2\right)^{\alpha }}{\Omega }\right)^k}{k! \  l!} \\
    - & \sum _{k,l=0}^{\infty} \frac{i^{N+1} (-1)^{k+1} \Gamma (l+k \alpha +\alpha  \mu ) \Gamma (l+N,\gamma ) \left(\frac{\mu  \left(2 \sigma ^2\right)^{\alpha }}{\Omega }\right)^k}{k! \ l! \ \Gamma (l+N)} 
\end{align}
\normalsize
Note that the first series in \eqref{eq: I2 split} is identical to $\mathcal{I}_1$; hence, they will cancel each other. 
Then, after minor simplifications, we finally obtain
\begin{align}
    \label{eq: PD Series}
    P_{\text{D}_{\mathcal{W}}} &= \frac{\alpha  \Psi^{\mu }}{\Gamma (\mu )} \sum _{k,l=0}^{\infty } \frac{\Gamma (l+N,\gamma ) \Gamma (l+k \alpha +\alpha  \mu ) \left(-\Psi\right)^k}{k! \   l! \ \Gamma (l+N)},
\end{align}
where $\Psi=\mu \left(2 \sigma ^2\right)^{\alpha }/ \Omega$. 
It is worth mentioning that \eqref{eq: PD Series} is also an original contribution of this work, enjoying a low computational burden as compared to \cite[Eq. 30]{Cui14}.

\section{Sample Numerical Results}
\label{eq: Sample Numerical Results}
\begin{algorithm}[t]
\DontPrintSemicolon
  \KwInput{$\tilde{\alpha}_n$, $\tilde{\Omega}_n$ and $N$}
  \textbf{Do:}  Compute the moments of $\eta$  and $\xi_{n}$ by using Eqs.~\eqref{eq: Moments eta} and~\eqref{eq: Moments xi}. \\
  \textbf{Do:} Solve numerically the parameters $\alpha$, $\mu$ and $\Omega$ by using Eqs.~\eqref{eq: Match 1}--\eqref{eq: Match 3}.\\
  \textbf{Do:} Apply Eq.~\eqref{eq: Final FOX} or, alternatively, Eq.~\eqref{eq: PD Series}.\\
\KwOutput{$P_{\text{D}_{\mathcal{W}}}$}
\caption{Computation of $P_{\text{D}_{\mathcal{W}}}$}
\label{alg: Algorithm PD}
\end{algorithm}
\begin{figure}[t]
\begin{center}
\includegraphics[trim={0cm 0cm 0cm 0cm},clip,scale=0.42]{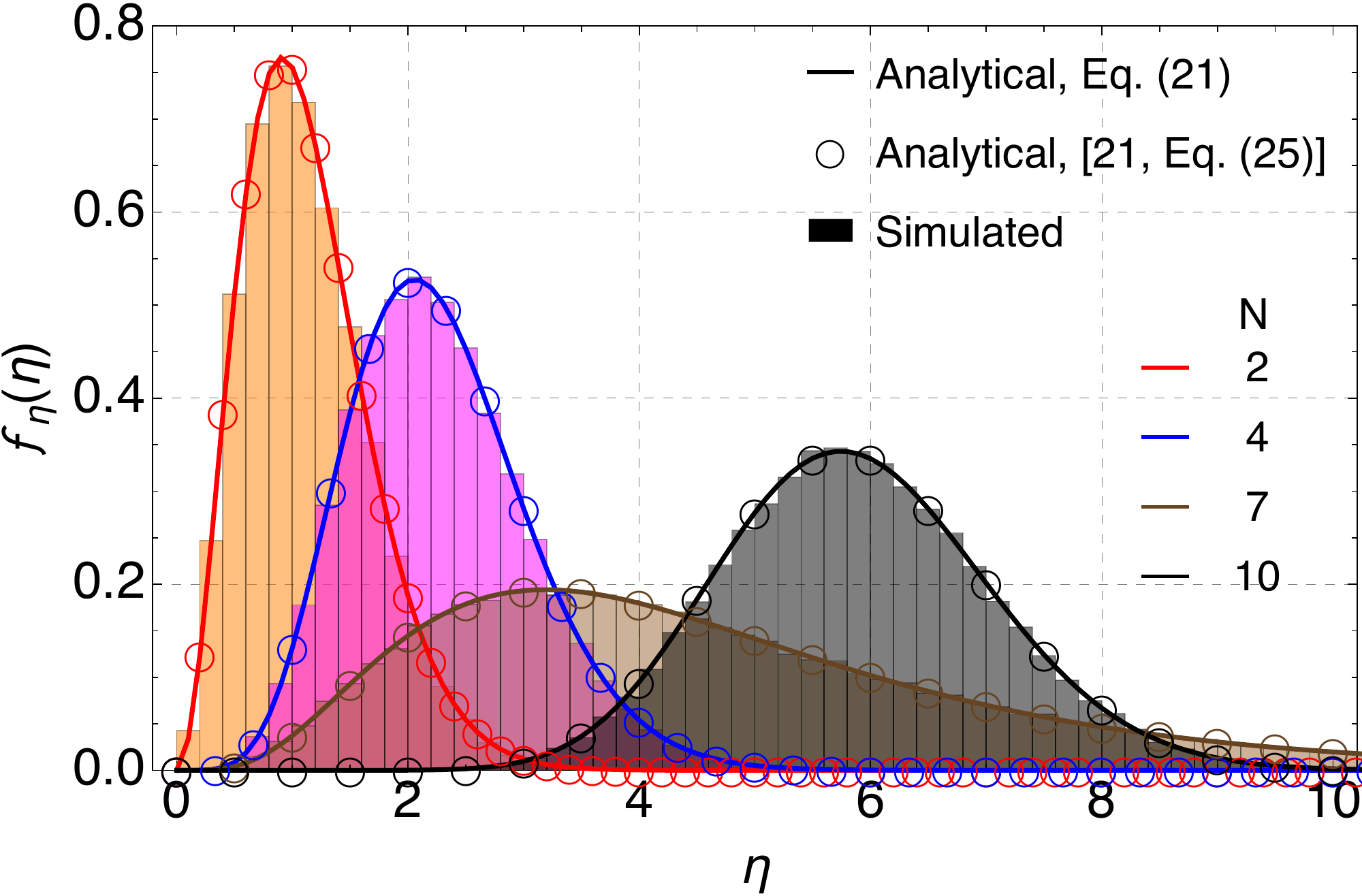}
\caption{PDF of $\eta$ for $\tilde{\Omega}_n=1/2$, $\tilde{\alpha}_n=3/2$, and different values of $N$.}
\label{fig: PDF Z}
\end{center} 
\end{figure}
\begin{figure}[t]
\begin{center}
\includegraphics[trim={0cm 0cm 0cm 0cm},clip,scale=0.42]{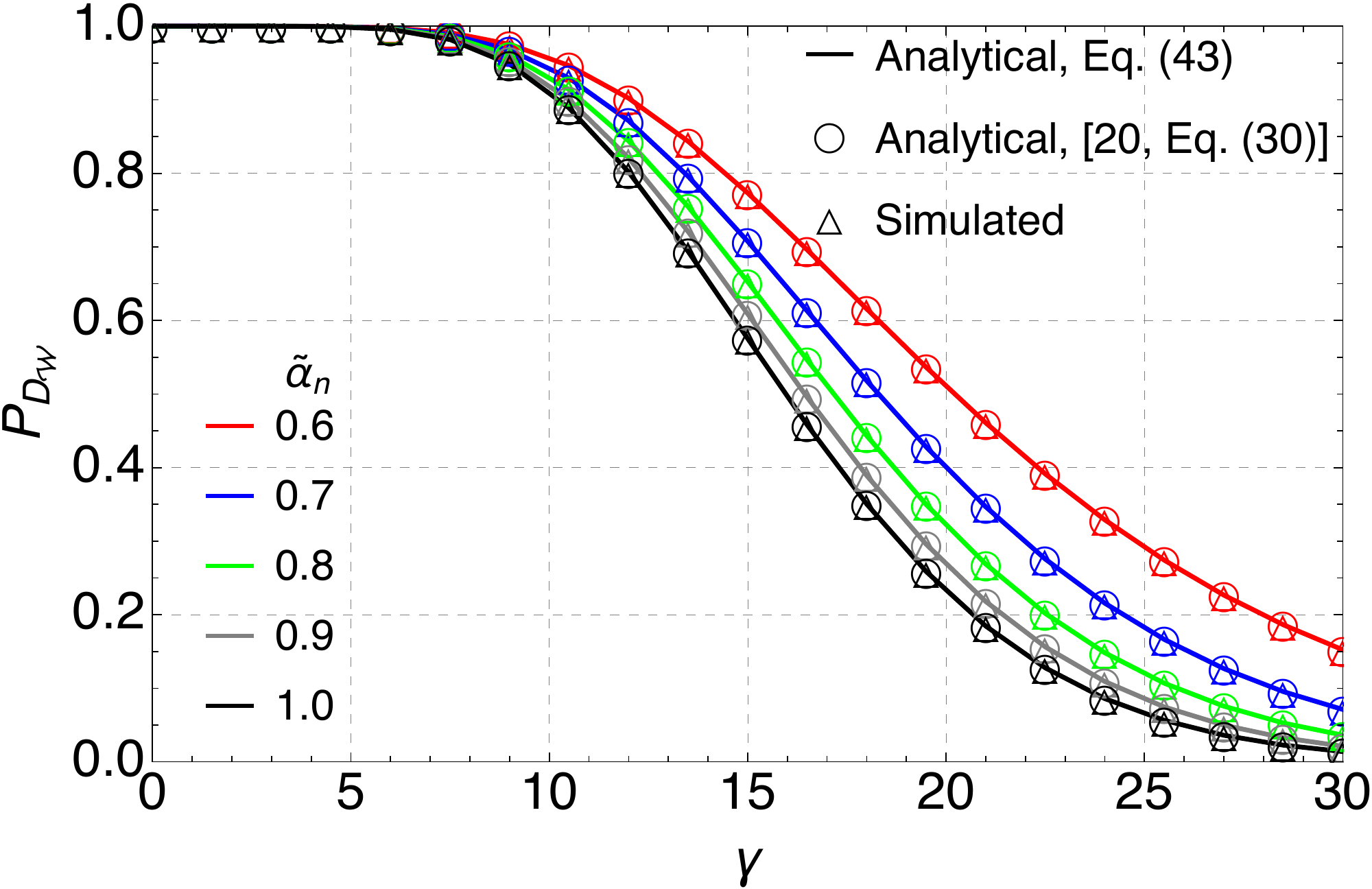}
\caption{$P_{\text{D}_{\mathcal{W}}}$ versus $\gamma$ for $\tilde{\Omega}_n=13/10$, $\sigma_0^2=1$, $N=10$ and different values of $\tilde{\alpha}_n$.}
\label{fig: PDvsLimiar alpha}
\end{center} 
\end{figure}
\begin{figure}[t]
\begin{center}
\includegraphics[trim={0cm 0cm 0cm 0cm},clip,scale=0.42]{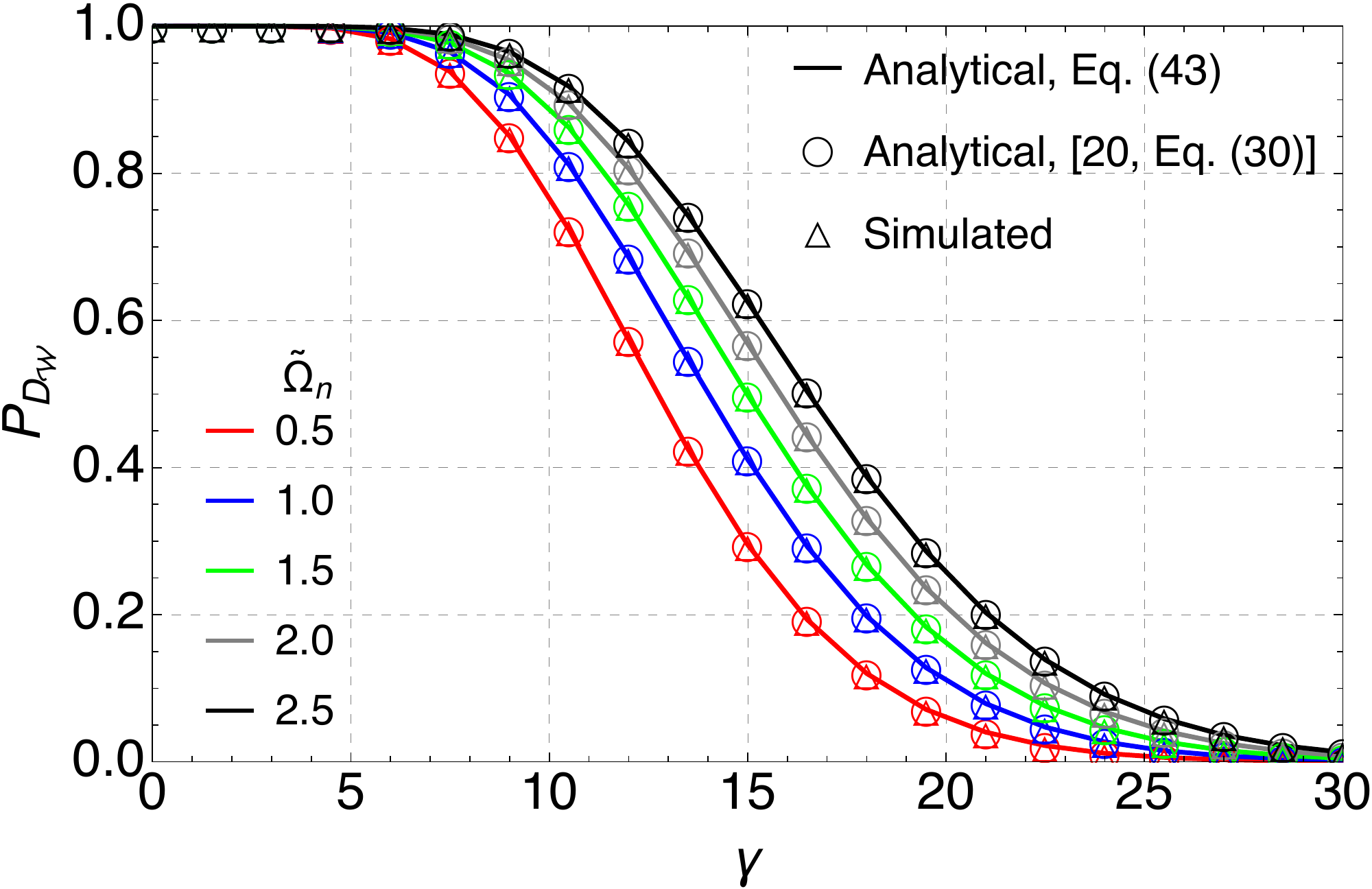}
\caption{$P_{\text{D}_{\mathcal{W}}}$ versus $\gamma$ for $\tilde{\alpha}_n=2$, $\sigma_0^2=1$, $N=10$ and different values of $\tilde{\Omega}_n$.}
\label{fig: PDvsLimiar Omega}
\end{center} 
\end{figure}
\begin{figure}[t]
\begin{center}
\includegraphics[trim={0cm 0cm 0cm 0cm},clip,scale=0.42]{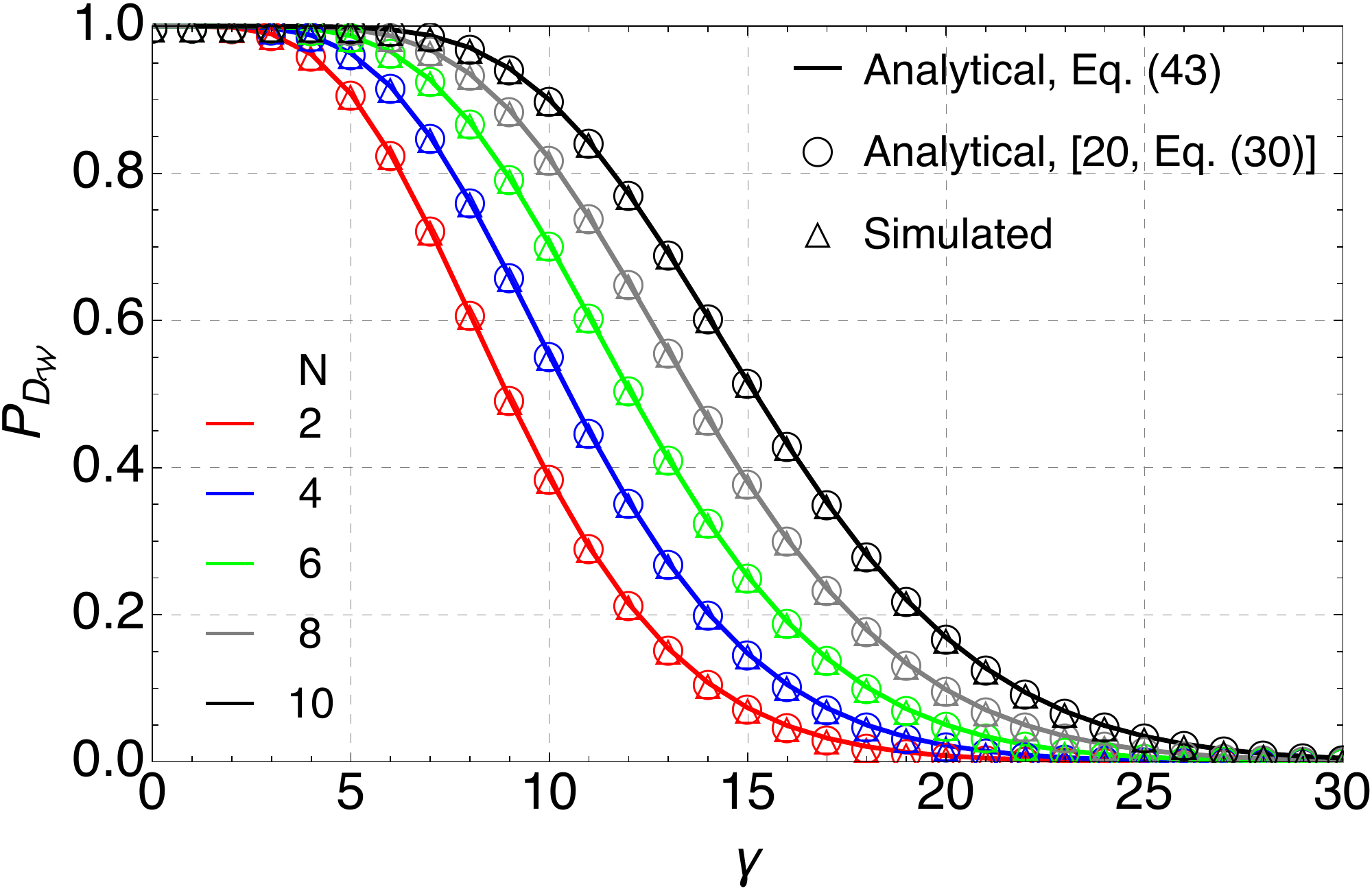}
\caption{$P_{\text{D}_{\mathcal{W}}}$ versus $\gamma$ for $\tilde{\alpha}_n=3$, $\tilde{\Omega}_n=2$, $\sigma_0^2=1$ and different values of $N$.}
\label{fig: PDvsLimiar N}
\end{center} 
\end{figure}
\begin{figure}[t]
\begin{center}
\includegraphics[trim={0cm 0cm 0cm 0cm},clip,scale=0.42]{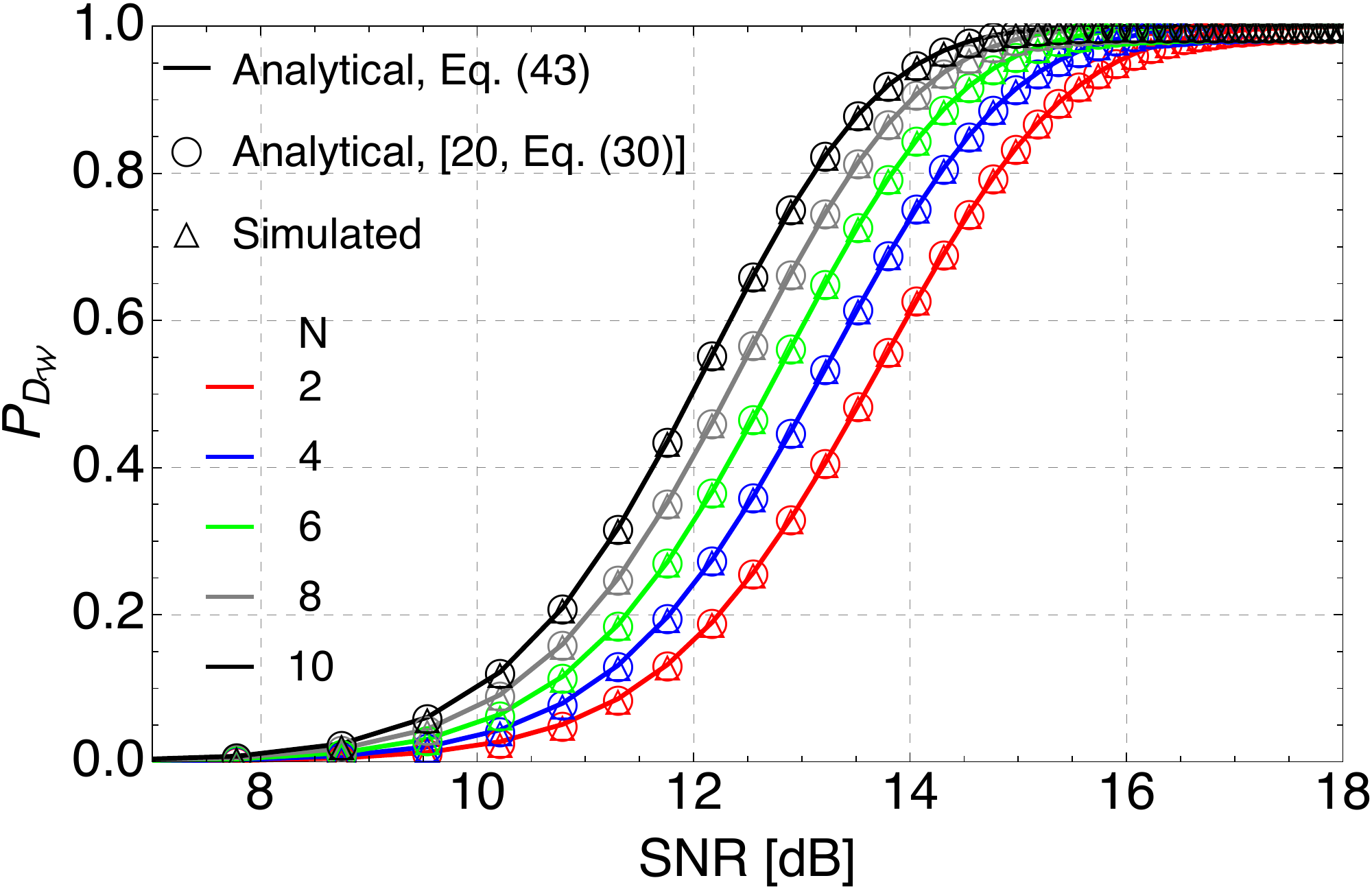}
\caption{$P_{\text{D}_{\mathcal{W}}}$ versus SNR for different values of $N$.}
\label{fig: SNRvsN}
\end{center} 
\end{figure}
\begin{figure}[t]
\begin{center}
\includegraphics[trim={0cm 0cm 0cm 0cm},clip,scale=0.42]{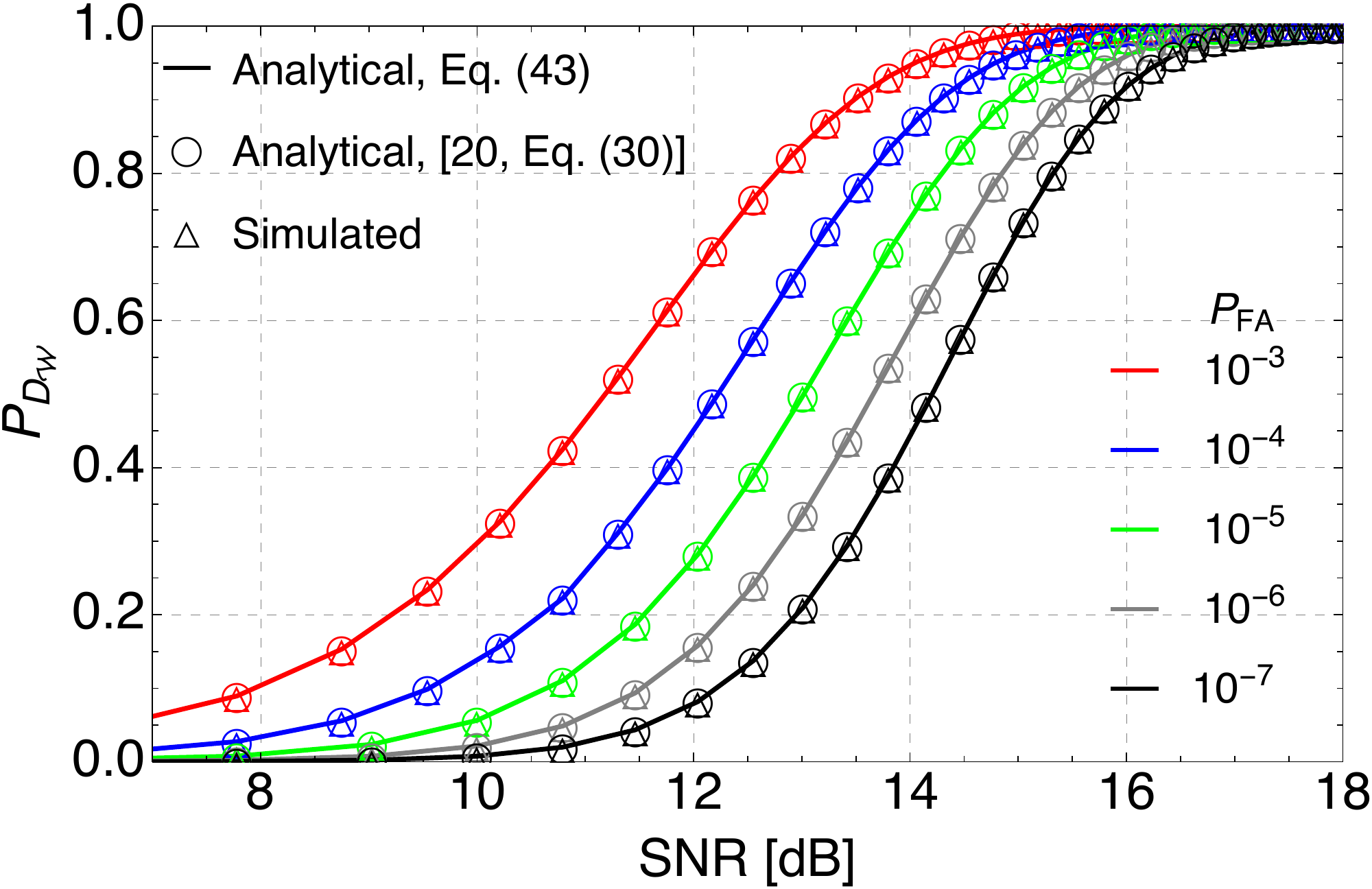}
\caption{$P_{\text{D}_{\mathcal{W}}}$ versus SNR for different values $P_{\text{FA}}$.}
\label{fig: SNRvsPFA}
\end{center} 
\end{figure}
\begin{table*}[t]
\centering
\caption{Efficiency of~\eqref{eq: PD Series} as compared to~\cite[Eq. (30)]{Cui14}.}
\label{tab: Efficiency}
\begin{tabular}{lccccc}
\hline \hline
\multicolumn{1}{c}{Parameter Settings }  & $P_{\text{D}_{\mathcal{W}}}$ $[\%]$ &$\mathcal{T}$  & \begin{tabular}[c]{@{}c@{}}Computation Time \\ for~\cite[Eq. (30)]{Cui14} [$s$] \end{tabular} &\begin{tabular}[c]{@{}c@{}}Computation Time \\ for Eq.~\eqref{eq: PD Series} [$s$] \end{tabular}    & \begin{tabular}[c]{@{}c@{}}Time\\ saving [\%] \end{tabular}\\ \hline
$N=3$, $\tilde{\alpha}_n=1/2$, $\tilde{\mu}_n=3/2$, $\tilde{\Omega}_n=2$, $\sigma^2=1$, $\gamma=3$ & $69.1485$ & $15.34$ $\times 10^{-4}$  & $1341.64$  &$27.2091$   &  $97.9721$\\ \hline
$N=3$, $\tilde{\alpha}_n=1$, $\tilde{\mu}_n=1$, $\tilde{\Omega}_n=2$, $\sigma^2=1$, $\gamma=2$ & $83.1095$  & $11.14$ $\times 10^{-4}$  & $1543.54$  & $45.0135$ & $97.0837$ \\ \hline
$N=3$, $\tilde{\alpha}_n=1/2$, $\tilde{\mu}_n=2$, $\tilde{\Omega}_n=5$, $\sigma^2=1$, $\gamma=3$ & $88.7412$ & $32.44$ $\times 10^{-4}$  &  $1711.14$   & $43.9077$   & $97.4341$ \\ \hline
$N=5$, $\tilde{\alpha}_n=1/2$, $\tilde{\mu}_n=3/2$, $\tilde{\Omega}_n=2$, $\sigma^2=1$, $\gamma=2$ & $97.6474$ & $49.13$ $\times 10^{-4}$   & $1579.19$  & $26.4729$  &  $98.3236$ \\
\hline
$N=5$, $\tilde{\alpha}_n=1/2$, $\tilde{\mu}_n=1$, $\tilde{\Omega}_n=2$, $\sigma^2=1$, $\gamma=2$ & $90.2578$ & $88.12$ $\times 10^{-4}$  & $1613.22$  & $46.2803$  &  $97.1312$ \\
\hline
$N=5$, $\tilde{\alpha}_n=1$, $\tilde{\mu}_n=1/2$, $\tilde{\Omega}_n=5$, $\sigma^2=1$, $\gamma=2$ & $97.5109$ & $57.58$ $\times 10^{-4}$  & $1887.91$  & $46.9172$  & $97.5149$ \\ \hline
$N=5$, $\tilde{\alpha}_n=1/3$, $\tilde{\mu}_n=3$, $\tilde{\Omega}_n=2$, $\sigma^2=1$, $\gamma=2$  & $92.0891$ & $92.33$ $\times 10^{-4}$ & $1787.32$  & $48.5396$  &  $97.2842$ \\
\hline
$N=6$, $\tilde{\alpha}_n=1/4$, $\tilde{\mu}_n=3$, $\tilde{\Omega}_n=1$, $\sigma^2=1$, $\gamma=1$ & $99.9091$ & $19.82$ $\times 10^{-4}$  & $1923.67$  & $46.0083$ & $97.6083$ \\ \hline
$N=6$, $\tilde{\alpha}_n=1/5$, $\tilde{\mu}_n=2$, $\tilde{\Omega}_n=1/2$, $\sigma^2=1$, $\gamma=1$ & $99.9387$  & $22.32$ $\times 10^{-4}$  & $1829.53$  & $48.4967$  & $97.3492$ \\ \hline
$N=5$, $\tilde{\alpha}_n=1/2$, $\tilde{\mu}_n=3/2$, $\tilde{\Omega}_n=2$, $\sigma^2=1$, $\gamma=3$ & $99.9413$ & $67.12$ $\times 10^{-4}$  & $1876.83$   & $51.5077$  & $97.2556$ \\
\hline  \hline
\end{tabular}
\end{table*}
In this section, we corroborate the validity of our expressions through Monte-Carlo simulations and numerical integration.\footnote{The number of realizations for Monte-Carlo simulations was set to $10^{6}$.} In addition, we illustrate the accuracy and low computational burden of~\eqref{eq: PD Series}.
Here, $P_{\text{D}_{\mathcal{W}}}$ was computed by performing the three steps described in Algorithm~\ref{alg: Algorithm PD}.

Fig. \ref{fig: PDF Z} shows the analytical and simulated PDF of $\eta$. The PDF parameters have been selected to show the wide range of shapes that the PDF can exhibit. 
Note the perfect agreement between the approximation proposed in \cite{Filho06}, the exact formulation in \cite{Yilmaz09conf}, and Monte-Carlo simulations. 

Figs.~\ref{fig: PDvsLimiar alpha}--\ref{fig: PDvsLimiar N} show $P_{\text{D}_{\mathcal{W}}}$ versus $\gamma$ by varying $\tilde{\alpha}_n$, $\tilde{\Omega}_n$ and $N$.
In all cases, observe the outstanding accuracy between our derived expressions and \cite[Eq. (30)]{Cui14}.
Also, note that the detection performance improves as $\tilde{\Omega}_n$ and $N$ increase, as expected. Similarly, the detection improves as $\tilde{\alpha}_n$ is reduced. 

Fig. \ref{fig: SNRvsN} shows  $P_{\text{D}_{\mathcal{W}}}$ versus SNR for different values of $N$. Note that for a fixed SNR, the higher the number of antennas, the better the radar detection. For example, given a $\text{SNR}=14$ dB, we obtain $P_{\text{D}_{\mathcal{W}}}=0.61,0.73,0.83,0.91,0.94$ for $N=2,4,6,8,10$, respectively.

Fig. \ref{fig: SNRvsPFA} shows  $P_{\text{D}_{\mathcal{W}}}$ versus SNR for different values of $P_{\text{FA}}$. Note that the radar performance improves as $P_{\text{FA}}$ is increased.  
This fundamental trade-off means that if $P_{\text{FA}}$ is reduced, $P_{\text{D}_{\mathcal{W}}}$ decreases as well.
For example, given a $\text{SNR}=14$ dB,  we obtain $P_{\text{D}_{\mathcal{W}}}=0.48,0.59,0.76,0.86,0.95$ for $P_{\text{FA}}=10^{-7},10^{-6},10^{-5},10^{-4},10^{-3}$, respectively.

Now, we evaluate the efficiency of~\eqref{eq: PD Series}. In order to so, we define 10 parameter settings, each with its corresponding $P_{\text{D}_{\mathcal{W}}}$, truncation error and the associated time saving to achieve the same accuracy goal imposed to~\cite[Eq. (6)]{Cui14}, say, around~$10^{-4}$, as shown in Table~\ref{tab: Efficiency}.
The truncation error is expressed~as
\begin{equation}
    \mathcal{T} = | P_{\text{D}_{\mathcal{W}}} - \overline{P_{\text{D}_{\mathcal{W}}}}|,
\end{equation}
where $\overline{P_{\text{D}_{\mathcal{W}}}}$ is the probability of detection obtained via the numerical integration of~\cite[Eq. (6)]{Cui14}. 
Observe that across all scenarios, the computation time dropped dramatically, showing an impressive reduction above $97$\%. 
Moreover,~\eqref{eq: PD Series} requires less than 275 terms to guarantee a truncation error of about $10^{-4}$.

\section{Conclusions}
\label{sec: Conclusions}
In this paper, we derived a highly accurate approximation for the PD of a non-coherent detector operating with Weibull fluctuation targets.
This approximation is given in terms of both a closed-form expression and a fast converging series. 
Numerical results and Monte-Carlo simulations corroborated the validity of our expressions, and showed the accuracy and fast rate of convergence of our results.
For instance, our series representation proved to be more tractable and faster than ~\cite[Eq. (30)]{Cui14}, showing an impressive reduction in computation time (above $97$\%) and in the required number of terms (less than 275 terms) to guarantee a truncation error of about $10^{-4}$.
The contributions derived herein allow us to reduce the computational burden that demands the PD evaluation. Moreover, they can be quickly executed on an ordinary desktop computer, serving as a useful tool for radar designers.

\newpage
\begin{appendices}
\section{Mathematica Implementation for the  Trivariate Fox H-Function}
\label{sec: Mathematica Implementation}
\lstset{
	tabsize=4,
	frame=single,
	language=mathematica,
	basicstyle=\scriptsize\ttfamily,
	keywordstyle=\color{black},
	backgroundcolor=\color{white},
	commentstyle=\color{magenta},
	showstringspaces=false,
	emph={
		If, Newton,Newton_, x_, delta_, delta, D_, D,beta_,beta, B_, M_, M, s, s_,B
	},emphstyle={\color{olive}},
	emph={[2] H, ClearAll, Remove, H_, Ns, value
	},emphstyle={[2]\color{blue}},
	emph={[3] L, L_, Result, ""Result"",UpP, LoP, Theta, R1, T1, R2, T2, k, R1, T1, R2, T2, m, n,kernel, S, W, t},emphstyle={[3]\color{magenta}}
	}

\begin{CenteredBox}
\begin{lstlisting}[caption={},linewidth=6.8cm, label=code:label]
ClearAll["Global`*"]; Remove[s];
H[x_, delta_, D_,beta_, B_]
 := Module[{UpP,LoP,Theta,R1,T1,R2,T2,m,n},
 L=Length[Transpose[D]]; 
 m=Length[D]; (*Number of Gamma functions in
 the numerator*)
 n=Length[B]; (*Number of Gamma functions in
 the denominator*)
 S=Table[Subscript[s,i],{i,1,L}]; (*s is the 
 vector containing the number of branches, in 
 our case s=[s_1,s_2,s_3]*)
 UpP=Product[Gamma[delta[[1,j]]+Sum[D[[j,k]]
     S[[k]],{k,1, L}]], {j,1,m}];
 LoP=Product[Gamma[beta[[1,j]]+Sum[B[[j,k]]
     S[[k]],{k,1,L}]],{j,1,n}];
 Theta=UpP/LoP (*Theta computes Eq. (2)*);
 W=50; (*Limit for the complex integration.
 Increase "MaxRecursion" for large W.*)
 T=Table[delta[[1,j]]+Sum[D[[j,k]]
 S[[k]],{k,1,L}]>0,{j,1,m}]; 
 (*Generation of the restriction table*)
 limit1 = -1(*Minimum limit of recursion*);
 limit2 = 1(*Maximum limit of recursion*);
 spacing = 1/2;
 T1=T/.{Subscript[s,1]->eps1,Subscript[s,2]
   ->eps2,Subscript[s,3]->eps3};
 Do[eps1=i;Do[eps2=j;Do[eps3=k;
   flag1=If[Total[Boole[T1]]==m,1,0];
   If[Total[T1]==m,Break[]],
   {k,limit1,limit2,spacing}];
   If[flag1==1,Break[]],
   {j,limit1,limit2,spacing}];
   If[flag1==1,Break[]],
   {i,limit1,limit2,spacing}];
   (*Find eps1, eps2 and eps3, nedded to
   separate the poles of left in Eq. (2), 
   from those of the right.*)
 kernel=Theta(x[[1]])^(-S[[1]])(x[[2]])
 ^(-S[[2]]) (x[[3]])^(-S[[3]])
 /.{S[[1]]->s1,S[[2]]->s2,S[[3]]->s3}; 
 (*Construction of the integratiion kernel*)
 Result= N[1/(2*Pi*I)^2 NIntegrate[kernel,
  {s1,-W-eps1*I,1/2-eps1*I},
  {s2,-W-eps2*I,1/2-eps2*I},
  {s3,-W-eps3*I,1/2-eps3*I},
  Method->{"GlobalAdaptive",
  Method->{"GaussKronrodRule"},
 "MaxErrorIncreases"->1000},
  MaxRecursion->20,AccuracyGoal->5,
  WorkingPrecision->20],20];
 Print[""Result""]];
\end{lstlisting}
\end{CenteredBox}
\end{appendices}
\bibliographystyle{IEEEtran}
\bibliography{MainArticle}
\end{document}